\documentclass{jfm}

\usepackage{natbib}
\usepackage{amsbsy}
\usepackage{psfrag}
\usepackage{amsmath}
\usepackage{bm}
\usepackage{soul}
\usepackage{verbatim}
\usepackage[dvipsnames]{xcolor}
\usepackage{graphicx}\usepackage{lineno}
\usepackage[export]{adjustbox}
\usepackage{tikz}
\usepackage{booktabs}
\usepackage{multirow}
\usetikzlibrary{shapes}
\definecolor{blueblue}{RGB}{0,114,190}
\definecolor{greengreen}{rgb}{0.16, 0.67, 0.53}
\definecolor{brickred}{rgb}{0.7960, 0.2550, 0.3290}

\usepackage{float}
\usepackage{placeins}

\newcommand{\subfigimgtwo}[3][,]{%
  \setbox1=\hbox{\includegraphics[#1]{#3}}
  \leavevmode\rlap{\usebox1}
  \rlap{\hspace*{-5pt}\raisebox{\dimexpr\ht1-1\baselineskip}{#2}}
  \phantom{\usebox1}
}
\newcommand{\subfigimgthree}[3][,]{%
  \setbox1=\hbox{\includegraphics[#1]{#3}}
  \leavevmode\rlap{\usebox1}
  \rlap{\hspace*{-10pt}\raisebox{\dimexpr\ht1-1\baselineskip}{#2}}
  \phantom{\usebox1}
}

\usepackage{url}
\usepackage[colorlinks=true,
             linkcolor=blue,
             urlcolor=blue,
             citecolor=blue]{hyperref}

\shorttitle{Actuator disk induction, thrust, and power in yaw misalignment} 
\shortauthor{K. S. Heck, H. M. Johlas and M. F. Howland} 
\title{Modeling the induction, thrust, and power of a yaw misaligned actuator disk}
\author
{
K. S. Heck\aff{1}, H. M. Johlas\aff{1}, and M. F. Howland\aff{1}\corresp{\email{mhowland@mit.edu}}
}
\affiliation
{
\aff{1}
Civil and Environmental Engineering, Massachusetts Institute of Technology, \\
Cambridge, MA 02139, USA
}

\begin{document}

\maketitle

\begin{abstract}
Collective wind farm flow control, where wind turbines are operated in an individually suboptimal strategy to benefit the aggregate farm, has demonstrated potential to reduce wake interactions and increase farm energy production.
However, existing wake models used for flow control often estimate the thrust and power of yaw misaligned turbines using simplified empirical expressions which require expensive calibration data and do not accurately extrapolate between turbine models.
The thrust, wake velocity deficit, wake deflection, and power of a yawed wind turbine depend on its induced velocity.
Here, we extend classical one-dimensional momentum theory to model the induction of a yaw misaligned actuator disk.
Analytical expressions for the induction, thrust, initial wake velocities, and power are developed as a function of the yaw angle and thrust coefficient.
The analytical model is validated against large eddy simulations of a yawed actuator disk.
Because the induction depends on the yaw and thrust coefficient, the power generated by a yawed actuator disk will always be greater than a $\cos^3(\gamma)$ model suggests, where $\gamma$ is yaw.
The power lost by yaw depends on the thrust coefficient.
An analytical expression for the thrust coefficient that maximizes power, depending on the yaw, is developed and validated.
Finally, using the developed induction model as an initial condition for a turbulent far-wake model, we demonstrate how combining wake steering and thrust (induction) control can increase array power, compared to either independent steering or induction control, due to the joint dependence of the induction on the thrust coefficient and yaw angle.
\end{abstract}


\section{Introduction}

Wake interactions between individual horizontal axis wind turbines can reduce wind farm energy production by $10$--$20\%$ \cite[]{barthelmie2009modelling}.
Utility-scale wind turbines are controlled to maximize individual power production, rather than collective wind farm production \cite[]{boersma2017tutorial}.
Individual operation entails aligning each wind turbine in the farm with the incoming wind direction.
In contrast, wake steering, where individual wind turbines are intentionally yaw misaligned with respect to the incident wind direction, has emerged as a promising strategy to reduce wake interactions and increase collective wind farm power production \cite[e.g.][]{gebraad2016wind, kheirabadi2019quantitative, bastankhah2019wind, zong2021experimental, howland2022optimal}.
Maximizing collective wind farm power production through wake steering control generally involves a trade-off between the power lost by the yaw misaligned turbines and the power gained by the downwind waked turbines, compared to standard individual control \cite[e.g.][]{fleming2015simulation}.
Since the power-maximizing yaw misalignment angles for wake steering control are primarily estimated using simplified, analytical flow models \cite[]{gebraad2016wind, fleming2019initial, howland2022collective}, it is important to accurately model the dependence of wind turbine power production and wake velocities on the yaw misalignment angle.

Wind turbine power production generally decreases as a function of an increasing yaw misalignment ($\gamma$) magnitude since the component of the wind velocity which is perpendicular to the rotor decreases.
Textbook materials instruct that the power production of a yawed wind turbine will decrease following $\cos^3(\gamma)$ \cite[]{burton2011wind}. 
This estimate is based on the application of classical one-dimensional momentum theory with an incoming axial freestream wind speed of $u_\infty \cdot \cos(\gamma)$ perpendicular to the rotor.
However, wind turbines extract power from the winds at the rotor. 
The wind at the rotor is affected by the velocity induced by the wind turbine.
Since the induction depends on the wind turbine thrust force and the thrust force will decrease in yaw misalignment, the induction will depend on the yaw misalignment.
The $\cos^3(\gamma)$ model neglects the dependence of the induction on the yaw misalignment \cite[]{micallef2016review}.
Given the error incurred by the $\cos^3(\gamma)$ model, most analytical wind farm power models assume that the power of a yaw misaligned wind turbine follows $P_r(\gamma) = P(\gamma)/P(\gamma=0)=\cos^{P_p}(\gamma)$, where $P_p$ is an empirical, turbine-specific factor that needs to be tuned using experimental data \cite[]{dahlberg2005research, gebraad2016wind}.
However, such experiments are costly, since they require sustained operation of utility-scale wind turbines in suboptimal yaw misalignment angles \cite[]{howland2020influence}.
Further, the wide spread in $P_p$ values reported in the literature, typically between $1$\textless$P_p$\textless$3$, suggests that the cosine model is not universal to different turbine models \cite[]{dahlberg2005research,schreiber2017verification, liew2020analytical, howland2020influence}.
Accurate analytical predictions of $P_r(\gamma)$ remain an outstanding challenge \cite[]{hur2019review} -- as a starting point, in this study, we focus on analytical predictions of the induction and power production of yawed actuator disks.

Through analysis of an autogyro aircraft, \cite{glauert1926general} developed an equation for the area-averaged induction and the coefficient of power as a function of the yaw misalignment $\gamma$.
\cite{glauert1926general} also identified that the induction of a yawed actuator disk varies over the rotor area about its mean value -- this finding has been replicated in other actuator disk simulations and models (see review by \cite{hur2019review}).
Glauert's yawed actuator disk momentum theory is commonly used in blade element momentum (BEM) models of rotational wind turbine aerodynamics \cite[see e.g. review by][]{micallef2016review}.
Using the Bernoulli equation, \cite{shapiro2018modelling} proposed an equation for the dependence of the axial induction factor on the yaw misalignment of an actuator disk.
\cite{speakman2021wake} used the axial induction equation proposed by \cite{shapiro2018modelling} to model $P_r(\gamma)$ for a simulation with a thrust coefficient of $0.75$, which yielded improved power predictions compared to the $\cos^3(\gamma)$ model, but higher predictive error than a tuned $\cos^{P_p}(\gamma)$ with $P_p$ set to $1.88$.

Beyond modeling the power-yaw relationship (i.e. $P_r(\gamma)$), modeling the inviscid near-rotor wake region of a yawed actuator disk is important since inviscid models are often used as an initial condition for turbulent wake models which are used to predict wind farm power production \cite[]{frandsen2006analytical, bastankhah2016experimental, shapiro2018modelling}.
Therefore, it is equally important to accurately model the induction and the streamwise and spanwise velocity deficits at the outlet of the inviscid near-wake region for a yawed actuator disk.

Finally, a parallel line of research to wake steering has investigated methods for axial induction flow control, where individual wind turbines reduce the magnitude of their wind speed wake deficits by decreasing the thrust force \cite[]{annoni2016analysis}.
A promising flow control methodology combines wake steering and induction control \cite[]{munters2018dynamic} -- for such combined control, it is important to model the joint effect of the yaw misalignment and the wind turbine thrust coefficient on the power and wake deficit.

In this study, classical, inviscid momentum theory is extended to the yaw misaligned actuator disk.
Analytical expressions are developed for the rotor normal induction, the streamwise velocity deficit, the spanwise velocity deficit, the thrust, and the power production of an actuator disk as a function of yaw misalignment.
In \S\ref{sec:model}, a model is proposed based on a combination of momentum conservation, mass conservation, and the Bernoulli equation.
The model is validated against large eddy simulations (LES) of a yawed actuator disk.
The numerical setup of the LES is given in \S\ref{sec:les} and results are provided in \S\ref{sec:results}.
The model is validated against the LES in \S\ref{sec:validation}.
The dependence of the induction, velocity deficits, and the power on the wind turbine thrust coefficient is presented in \S\ref{sec:optimal_ctp}.
Further, in \S\ref{sec:optimal_ctp}, the model is optimized to find the thrust coefficient which maximizes power for each value of the yaw misalignment angle.
In \S\ref{sec:steering}, the induction model is used as an initial condition for a turbulent far-wake model. 
The implications of the developed induction-yaw model on quasi-steady wake steering and induction control are presented and discussed. 
Conclusions are provided in \S\ref{sec:conclusions}.

\begin{figure}
    \centering
    \includegraphics[width=\linewidth]{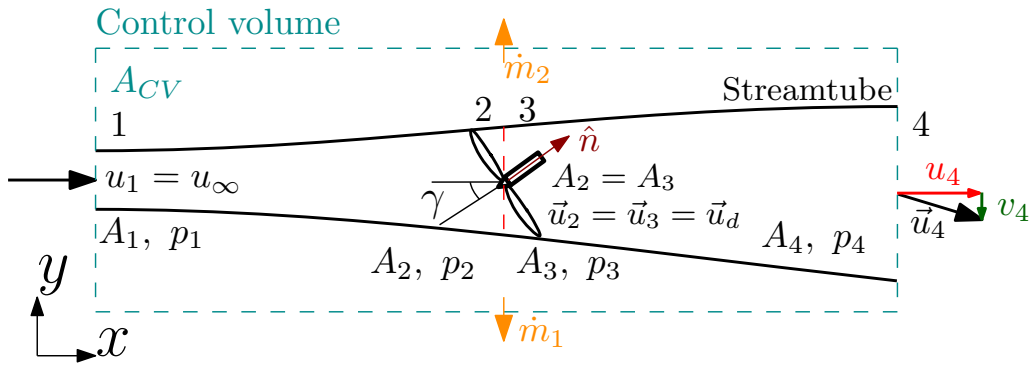}
    \caption{Control volume for the yawed actuator disk analysis.
    The streamwise and spanwise directions are $x$ and $y$, respectively.
    The actuator disk modeled wind turbine is yaw misaligned at angle $\gamma$, where positive yaw misalignment is a counter-clockwise rotation viewed from above.
    As in classical momentum theory, we consider four stations for the analysis, and the flow variables are labeled with the corresponding station as subscript numbers.
    The cross-sectional areas, streamwise velocities, spanwise velocities, pressures, and mass flow rates are denoted as $A$, $u$, $v$, $p$, and $\dot{m}$, respectively.
    The unit vector normal to the yawed wind actuator disk is shown as $\hat{n}$.}
    \label{fig:cv}
\end{figure}

\section{Yawed actuator disk momentum theory}
\label{sec:model}



Our goal is to model the induction, thrust, wake deficit and deflection, and the power production of a yaw misaligned actuator disk.
For the following analysis, we assume that the flow is inviscid and frictionless. 
We assume that the velocity is continuous across the actuator disk, including both the streamwise and spanwise velocities, and that the pressure recovers to the incident freestream pressure away from the actuator disk.
We note that the pressure recovery assumption is only relevant to the Bernoulli equation and streamwise momentum analysis.
We do not apply this pressure recovery assumption to a lateral momentum balance, since it is well-known to introduce predictive error \cite[]{shapiro2018modelling} due to counter-rotating vortices in the wake of yawed turbines \cite[]{howland2016wake}.
We consider uniform inflow and an actuator disk model (ADM) representation of the wind turbine forcing \cite[]{calaf2010large,burton2011wind}.
The ADM is introduced in \S\ref{sec:adm}.
The lateral velocity is modeled following lifting line theory \cite[]{shapiro2018modelling} (\S\ref{sec:spanwise}).
The induction is modeled by combining the Bernoulli equation, conservation of mass, and momentum conservation to a control volume containing the yaw misaligned actuator disk (\S\ref{sec:cv_full}).
A schematic of the yaw misaligned actuator disk and the control volume is shown in Figure~\ref{fig:cv}.

In \S\ref{sec:cv_full}, we develop the equations to predict the induction, thrust, wake deficit and deflection, and the power production of a yaw misaligned actuator disk.
In \S\ref{sec:limit_2}, we consider a limiting case of the developed induction model where the outlet spanwise velocity $v_4$ is negligible compared to the outlet streamwise velocity $u_4$, $|v_4| \ll u_4$. 

\subsection{Actuator disk model}
\label{sec:adm}

The thrust force from an actuator disk on the surrounding flow depends on the freestream rotor-normal wind speed, $\vec{u}_\infty \cdot \hat{n}: $ 
\begin{equation}
    \vec{F}_{T,\mathrm{ideal}} = -\frac{1}{2} \rho C_T A_d (\vec{u}_{\infty} \cdot \hat{n})^2 \hat{n},
\end{equation}
where $\rho$ is the density of the incident air, $C_T$ is the coefficient of thrust, $A_d=\pi D^2/4$ is the area of the rotor disk where $D$ is the wind turbine rotor diameter, $\hat{n}$ is the unit normal vector perpendicular to the disk, and $\vec{u}_{\infty}$ is the freestream wind velocity vector \cite[]{sorensen2011aerodynamic}. 
Wind turbines produce thrust and power based on the wind velocity at the rotor, which has been modified by induction.
Thus, the empirical thrust coefficient $C_T$ depends on the induction.
Additionally, for wind farms in the atmospheric boundary layer, it may be challenging to estimate the value of the freestream reference wind speed $u_\infty$ due to wakes of upstream turbines or heterogeneity in the background flow field.
Instead, an ADM is used to model wind turbine forcing, where the thrust force scales with the rotor-normal wind speed at the disk, $\vec{u}_d \cdot \hat{n}$ rather than the freestream $\vec{u}_\infty \cdot \hat{n}$ \cite[]{calaf2010large}.
The ADM thrust force then depends on a modified thrust coefficient $C_T'$ and the disk velocity \cite[]{calaf2010large}: 
\begin{equation}
    \vec{F}_T = -\frac{1}{2} \rho C_T^{\prime} A_d (\vec{u}_{d} \cdot \hat{n})^2 \hat{n}.
\label{eq:ft_full_les}
\end{equation}
Equation~\eqref{eq:ft_full_les} is used in the ADM implementation in LES used for validation as well as the derivation of the analytical model.

Assuming that the freestream wind is uniform and aligned with the $x$-direction, the freestream wind vector is $\vec{u}_\infty = u_\infty \hat{\imath}+ 0 \hat{\jmath}$. 
However, the disk velocity may include a component in the $y$-direction for yaw-misaligned turbines, and so is generally $\vec{u}_d = u_{d}\hat{\imath} + v_{d}\hat{\jmath}$. 
The rotor normal induction factor $a_n$ for a rotor with yaw misalignment angle $\gamma$ is defined as
\begin{equation}
    a_n = 1 - \frac{ \vec{u}_d \cdot \hat{n} }{ u_{\infty} \cos(\gamma)}. 
    \label{eq:an}
\end{equation}
In the yaw-aligned case where $\hat{n}=\hat{\imath}$, the rotor normal induction factor $a_n$ reduces to the standard (streamwise) axial induction factor $a = 1 - u_d / u_\infty$.
The thrust force written in terms of the rotor normal induction factor is then
\begin{equation}    
    \vec{F}_T = -\frac{1}{2} \rho C_T^{\prime} A_d ( 1 - a_n)^2 \cos^2(\gamma) u_\infty^2 \left[ \cos{(\gamma)} \hat{\imath} + \sin{(\gamma)} \hat{\jmath} \right].
    \label{eq:ft_full}
\end{equation}
The power for the actuator disk is computed as $P = -\vec{F}_T \cdot \vec{u}_d$.

Rotational, utility-scale wind turbines produce a thrust force which depends on the disk velocity \cite[e.g.][]{burton2011wind,sorensen2011aerodynamic,howland2020influence} -- the disk velocity is lower than $u_\infty$ due to induction.
Similarly, the ADM produces a thrust force which is proportional to the disk velocity, which has been modified by induction.
The thrust force depends on the yaw misalignment for both utility-scale, rotational wind turbines and the ADM.
In the ADM, $C_T^{\prime}$ is a fixed input.
Therefore, the thrust force of the ADM depends on the yaw misalignment following $F_T(\gamma) \propto (1-a_n(\gamma))^2 \cos^2(\gamma)$.
However, since the rotor normal induction depends on the imposed thrust force $F_T$, and the thrust force decreases with an increasing magnitude of yaw misalignment, we hypothesize that the induction factor will depend on $\gamma$. 

We emphasize that the following analysis will prescribe an ADM-type forcing where $C_T^\prime$ is a fixed quantity which does not depend on the yaw misalignment (see Eq.~\eqref{eq:ft_full_les}).
For different wind turbine models, a different form of the thrust force $F_T$ may be appropriate (i.e. a different form than Eq.~\eqref{eq:ft_full_les}).
Specifically, $C_T^\prime$ may not be a fixed quantity.
In general for rotational turbines, the potential dependence of $C_T^\prime$ on the yaw misalignment will depend on the turbine control strategy (i.e. the blade pitch and torque control) and the wind conditions \cite[]{howland2020influence}.
For a different form of $F_T$, the quantitative model predictions would differ but the qualitative trends of the influence of yaw misalignment on the induction are expected to apply.
We further discuss this detail and the need for future work in \S\ref{sec:conclusions}.

\subsection{Lifting line spanwise velocity model with rotor normal induction depending on yaw}
\label{sec:spanwise}

Yaw misaligned wind turbines generate a counter-rotating vortex pair (CVP) which deflects and deforms the wake region into a curled wake shape \cite[]{howland2016wake, bastankhah2016experimental, fleming2018simulation, martinez2021curled}.
The CVP rotates about a low-pressure center.
Momentum balance approaches to predict the lateral velocity in the wake of a yaw misaligned actuator disk which neglect the influence of the lateral pressure gradient often exhibit predictive errors \cite[]{jimenez2010application, shapiro2018modelling}.
\cite{shapiro2018modelling} developed a model for the spanwise velocity downwind of a yaw misaligned actuator disk.
The approach uses Prandtl lifting line theory \cite[]{milne1973theoretical} to predict the spanwise velocity in the inviscid near-wake region downwind of the actuator disk.
The downwash produced by the lifting line theory was presumed to be the spanwise velocity in the outlet of the streamtube enclosing the yawed actuator disk.
The resulting model predicts the spanwise velocity disturbance $\delta v_0 = v_\infty - v_4 = \frac{1}{4} C_T u_\infty \cos^2(\gamma) \sin(\gamma)$.
The model exhibited excellent predictions of the circulation at the disk hub-height ($z=0$), defined as $\Gamma_0$ \cite[]{shapiro2018modelling}, over a range of yaw and thrust values. 
The spanwise velocity disturbance $\delta v_0$ was also compared to LES.
The predictions exhibited improved accuracy compared to previous models, but had a slight underprediction of $\delta v_0$ at high yaw misalignment angles, $|\gamma|>20^\circ$ \cite[]{shapiro2018modelling}. 

Following \S\ref{sec:adm}, we consider the Prandtl lifting line approach developed by \cite{shapiro2018modelling} applied to the ADM with a prescribed $C_T^\prime$, instead of a prescribed $C_T$.
The spanwise velocity disturbance is
\begin{equation}
\delta v_0 = v_\infty - v_4 = \frac{- \Gamma_0}{4R} = \frac{-\vec{F}_T \cdot \hat{\jmath}}{2 \rho u_\infty A_d} = \frac{1}{4} C_T^\prime u_\infty \sin(\gamma) \cos^2(\gamma) \left(1-a_n(\gamma)\right)^2.
\label{eq:shapiro_mod}
\end{equation}
Comparing Eq.~\eqref{eq:shapiro_mod} to the model proposed by \cite{shapiro2018modelling}, $C_T^\prime$ is the input fixed quantity and there is an additional non-linear dependence on $a_n(\gamma)$.
We note that \cite{shapiro2018modelling} identified the influence of the yaw misalignment on the induction, and accounted for it by plotting $\delta v_0$ against $C_T$ where the thrust coefficient was empirically estimated as $C_T = C_T^\prime \tilde{u}_{d}^2 / (u_\infty^2 \cos^2(\gamma))$, where $\tilde{u}_d$ was the disk velocity measured from the LES validation case.
In the following sections, we will develop a predictive model for $a_n(\gamma)$ which uses Eq.~\eqref{eq:shapiro_mod}.

\subsection{Model for the induction of a yaw misaligned actuator disk}
\label{sec:cv_full}
To model the induction, we first apply the Bernoulli equation from stations 1 to 2 and stations 3 to 4 within the streamtube, shown in Figure~\ref{fig:cv}:
\begin{equation} \begin{split}
    p_1 + \frac{1}{2}\rho ||\vec{u}_1||^2 &= p_2 + \frac{1}{2}\rho ||\vec{u}_2||^2 \\
    p_3 + \frac{1}{2}\rho ||\vec{u}_3||^2 &= p_4 + \frac{1}{2}\rho ||\vec{u}_4||^2,
    \label{eq:bernoulli_1}
\end{split} \end{equation}
where $||\vec{u}_4|| = \sqrt{u_4^2 + v_4^2}$.
We note that the outlet flow has nonzero components in the $x$ ($\hat{\imath}$) and $y$ ($\hat{\jmath}$) directions, denoted as $u_4$ and $v_4$, respectively (see Figure~\ref{fig:cv}).
Assuming that the pressure recovers to the freestream at station 4 ($p_1$ = $p_4 = p_\infty$) and that the velocity across the rotor disk is continuous ($\vec{u}_2 = \vec{u}_3 = \vec{u}_d$), Eqs.~\eqref{eq:bernoulli_1} can be combined and simplified to
\begin{equation}
    p_2 - p_3 = \frac{1}{2}\rho \left( ||\vec{u}_1||^2 - ||\vec{u}_4||^2 \right).
\label{eq:bernoulli}
\end{equation}
Substituting in $\vec{u}_1 = \vec{u}_\infty$, $\vec{u}_4=u_4\hat{\imath} + v_4\hat{\jmath}$, and $(p_2 - p_3) A_d = ||\vec{F}_{T}||$ with $\vec{F}_T$ given by Eq.~\eqref{eq:ft_full}, this becomes
\begin{equation}
    u_\infty^2 - u_4^2 - v_4^2 = C_T' (1-a_n)^2 \cos^2(\gamma) u_\infty^2.
\label{eq:bernoulli_result}
\end{equation}
Next, we apply mass conservation to the streamtube between stations 2 and 4, where $A_2=A_d$:
\begin{equation}
    \vec{u}_4 \cdot (A_4 \hat{\imath}) = \vec{u}_2 \cdot (A_2 \hat{n}).
\end{equation}
Substituting in $\vec{u}_2=\vec{u}_d$ and the definition of $a_n$ in Eq.~\eqref{eq:an}, $\vec{u}_d \cdot \hat{n} = (1-a_n)u_\infty \cos(\gamma)$, this simplifies to:
\begin{equation}
    u_4 A_4 = (1-a_n) u_\infty \cos(\gamma)  A_d.
\label{eq:streamtube}
\end{equation}
We then apply mass conservation to the two-dimensional control volume, assuming that the flow outside the disk streamtube is unperturbed at $u_\infty$:
\begin{equation} \begin{split}
    \dot{m}_{1} + \dot{m}_{2} &= 
    \rho \vec{u}_\infty \cdot \vec{A}_{CV} - 
    \rho \vec{u}_\infty \cdot (\vec{A}_{CV} - \vec{A}_4) - 
    \rho \vec{u}_4 \cdot \vec{A}_4 
    \\
    &= \rho A_4 (u_\infty - u_4),
\label{eq:mass_cons}
\end{split} \end{equation}
where $\mathrm{CV}$ denotes the control volume (Figure~\ref{fig:cv}).
Finally, we apply conservation of momentum to the control volume in the streamwise direction ($\hat{\imath}$), using the Reynolds transport theorem assuming steady-state flow:
\begin{equation}
     \rho \frac{\mathrm{D}u}{\mathrm{D} t} = \int_{\mathrm{CS}} \rho u \left(\vec{u}_{\mathrm{rel}} \cdot \mathrm{d}\vec{A}\right) = \vec{F}_T \cdot \hat{\imath} + p_1 A_{CV} - p_4 A_{CV},
\label{eq:mom_cons}
\end{equation}
where $\mathrm{CS}$ is the control surface.
By expanding the surface integral and combining terms, this momentum balance simplifies to
\begin{equation}
\vec{F}_T \cdot \hat{\imath} =  \rho u_4^2 A_4 - \rho u_\infty^2 A_4 + (\dot{m}_{1} + \dot{m}_{2}) u_\infty.
\label{eq:mom_cons_2}
\end{equation}
Substituting Eqs.~\eqref{eq:ft_full}, \eqref{eq:streamtube}, and \eqref{eq:mass_cons} into Eq.~\eqref{eq:mom_cons_2} and simplifying gives
\begin{equation} \begin{split}
     -\frac{1}{2} C_T^{\prime} u_\infty (1 - a_n) \cos^2(\gamma) &= u_4 - u_\infty. 
    \label{eq:cons_result}
\end{split} \end{equation}

Finally, we solve for $a_n$ in Eq.~\eqref{eq:bernoulli_result} from Bernoulli, $u_4/u_\infty$ in Eq.~\eqref{eq:cons_result} from conservation of mass and the streamwise momentum balance, and $v_4/u_\infty$ in Eq.~\eqref{eq:shapiro_mod} from the lifting line spanwise velocity deficit model, resulting in a coupled nonlinear system of three equations to solve for $a_n(\gamma)$, $u_4(\gamma)$, and $v_4(\gamma)$:
\begin{equation}
    \begin{cases}
    a_n(\gamma) = 1 - \cfrac{\sqrt{u_\infty^2-u_4(\gamma)^2-v_4(\gamma)^2}}{ \sqrt{C_T^\prime} u_\infty \cos(\gamma)}  & (a) \\
    \cfrac{u_4(\gamma)}{u_\infty} = 1 - \frac{1}{2} C_T^\prime \big(1-a_n(\gamma)\big) \cos^2(\gamma) & (b)\\
    \cfrac{v_4(\gamma)}{u_\infty} = -\frac{1}{4} C_T^\prime \big(1-a_n(\gamma)\big)^2 \sin(\gamma) \cos^2(\gamma)  & (c) 
    
    \end{cases}
    \label{eq:full_model}
\end{equation}
The system in Eq.~\eqref{eq:full_model} can be solved iteratively from an initial condition from standard, yaw-aligned momentum theory $a_n^0 = a = \frac{1}{2}(1-\sqrt{1-C_T})=C_T'/(C_T'+4)$ and typically converges in less than five iterations.
While the system of equations in Eq.~\eqref{eq:full_model} converges quickly, it does not permit a straightforward solution.
In \S\ref{sec:limit_2}, we examine a limiting case of the model where the outlet spanwise velocity is neglected in the Bernoulli equation, $|v_4| \ll u_4$.

With a solution for the normal induction factor $a_n(\gamma)$ from Eq.~\eqref{eq:full_model}, the power for a yaw misaligned actuator disk is modeled as
\begin{equation}
P(\gamma) = -\vec{F}_T \cdot \vec{u}_d = \frac{1}{2} \rho C_T^\prime A_d \big(1-a_n(\gamma)\big)^3 u_\infty^3 \cos^3(\gamma).    
\label{eq:power_ADM}
\end{equation}
As discussed in the introduction, the dependence of wind turbine power production on the yaw misalignment is often described by the power ratio $P_r(\gamma)$ \cite[e.g.][]{howland2020influence}.
The resulting model for the power ratio is
\begin{equation}
P_r(\gamma) = \frac{P(\gamma)}{P(\gamma=0)} = \left[\left(1+\frac{1}{4}C_T^\prime\right) (1-a_n(\gamma)) \cos(\gamma) \right]^3,
\label{eq:Pr_uniform_analytical_mom_full}
\end{equation}
 and the thrust ratio is 
\begin{equation}
T_r(\gamma) = \frac{F_T(\gamma)}{F_T(\gamma=0)} = \left[ \left(1+\frac{1}{4}C_T^\prime\right) (1-a_n(\gamma)) \cos(\gamma)  \right]^2.
\label{eq:Tr_uniform_analytical_mom_full}
\end{equation}

\subsection{Limiting case of CV analysis with $|v_4| \ll u_4$}
\label{sec:limit_2}

In this section, we consider the limiting case where the outlet spanwise velocity from the streamtube is significantly less than the outlet streamwise velocity, $|v_4| \ll u_4$.
Therefore, the outlet velocity is $||\vec{u}_4|| = u_4$.
Starting from Eq.~\eqref{eq:full_model}, the rotor normal induction is simplified as
\begin{equation}
a_n(\gamma) = \frac{C_T^\prime \cos^2(\gamma)}{4 + C_T^\prime \cos^2(\gamma)},
\label{eq:a_shapiro}
\end{equation}
which is also the induction factor reported by \cite{shapiro2018modelling}, who assumed that the spanwise velocity disturbance appeared infinitesimally downwind of the yawed actuator disk and that it was constant in the streamtube downwind.
The streamwise and spanwise velocities are
\begin{align}
\cfrac{u_4(\gamma)}{u_\infty} = \frac{4 - C_T^\prime \cos^2(\gamma)}{4 + C_T^\prime \cos^2(\gamma)},  && \cfrac{v_4(\gamma)}{u_\infty} = -\frac{4 C_T^\prime \sin(\gamma) \cos^2(\gamma)}{(4 + C_T^\prime \cos^2(\gamma))^2}. 
\end{align}
The streamwise outlet velocity $u_4(\gamma)$ can also be written in terms of the induction factor $a_n(\gamma)$ such that $u_4(\gamma) = u_\infty (1 - 2a_n(\gamma))$, where $a_n(\gamma)$ is given by Eq.~\eqref{eq:a_shapiro}. 
This is analogous to the outlet velocity from one-dimensional momentum $u_4(\gamma=0) = u_\infty(1-2a)$, where $a = a_n(\gamma=0)$ is again the standard axial induction factor. 

The power of the yawed actuator disk in this limiting case is
\begin{equation}
P(\gamma) = \frac{32 \rho A_d C_T^\prime \cos^3(\gamma) u_\infty^3}{(4+C_T^\prime \cos^2(\gamma))^3}.
\end{equation}
The power and thrust ratios in this limiting case are
\begin{align}
P_r(\gamma) = \left[\frac{(4+C_T^\prime) \cos(\gamma)}{4+C_T^\prime \cos^2(\gamma)}\right]^3, && T_r(\gamma)  = \left[\frac{(4+C_T^\prime) \cos(\gamma)}{4+C_T^\prime \cos^2(\gamma)}\right]^2.
\label{eq:Pr_uniform_analytical_mom_limit}
\end{align}
The power ratio model given by Eq.~\eqref{eq:Pr_uniform_analytical_mom_limit} was also reported by \cite{speakman2021wake}, who leveraged the streamwise induction model developed by \cite{shapiro2018modelling} (same as Eq.~\eqref{eq:a_shapiro}).

\section{Large eddy simulation numerical setup}
\label{sec:les}
Large eddy simulations are performed using an incompressible flow code Pad{\'e}Ops\footnote{https://github.com/FPAL-Stanford-University/PadeOps} \cite[]{ghate2017subfilter, howland2020influencea}.
Fourier collocation is used in the horizontally homogeneous directions and a sixth-order staggered compact finite difference scheme is used in the vertical direction \cite[]{nagarajan2003robust}.
Time advancement uses a fourth-order strong stability preserving (SSP) variant of Runge-Kutta scheme \cite[]{gottlieb2011strong} and the subgrid scale closure uses the sigma subfilter scale model \cite[]{nicoud2011using}.

The ADM is implemented with the regularization methodology introduced by \cite{calaf2010large} and further developed by \cite{shapiro2019filtered}. 
The ADM forcing depends on the prescribed input of $C_T^\prime$ (see Eq.~\eqref{eq:ft_full_les}), which is held constant for varying yaw misalignment angle. 
The discretized turbine thrust force $\vec{f}(\vec{x})$ is distributed in the computational domain ($\vec{x}$) through an indicator function $\mathcal{R}(\vec{x})$ as
\begin{equation}
\label{eq:LES_fx}
    \vec{f}(\vec{x}) = \vec{F}_T \mathcal{R}(\vec{x}). 
\end{equation}
The thrust force $\vec{F}_T$ is computed with Eq.~\eqref{eq:ft_full_les}, depending on the disk velocity $\vec{u}_d$.
The indicator function $\mathcal{R}(\vec{x})$ is constructed from a decomposition $\mathcal{R}(\vec{x}) = \mathcal{R}_1(x) \mathcal{R}_2(y, z)$ given by Eqs.~\eqref{eq:R1_func} and \eqref{eq:R2_func}
\begin{gather}
    \label{eq:R1_func}
    \mathcal{R}_1(x) = \frac{1}{2s} \left[
    \mathrm{erf}\left(\frac{\sqrt{6}}{\Delta}
    \left(x+\frac s2\right)\right) -
    \mathrm{erf}\left(\frac{\sqrt{6}}{\Delta}
    \left(x-\frac s2\right)\right)\right], 
    \\
    \label{eq:R2_func}
    \mathcal{R}_2(y, z) = \frac{4}{\pi D^2}\frac{6}{\pi\Delta^2} 
    \iint H\left(D/2 - \sqrt{y'^2+z'^2}\right) 
    \exp{\left(-6 \frac{(y-y')^2+(z-z')^2}{\Delta^2}\right)}\,dy'\,dz', 
\end{gather}
where $H(x)$ is the Heaviside function, $\mathrm{erf}(x)$ is the error function, $s$ is the ADM disk thickness, and $\Delta$ is the filter width. 
The disk velocity $\vec{u}_d$, used in the thrust force calculation Eq.~\eqref{eq:ft_full_les}, is calculated using the indicator function such that 
\begin{equation}
\label{eq:corr_ud}
    \vec{u}_d = M\iiint \mathcal{R}(\vec{x}) \vec{u}(\vec{x}) \,d^3\vec{x}, 
\end{equation}
where $\vec{u}(\vec{x})$ is the filtered velocity in the LES domain.
Depending on the numerical implementation of the indicator function, particularly the selection of filter width $\Delta$, the ADM can underestimate the induction and therefore overestimate power production \cite[]{munters2017optimal,shapiro2019filtered}. 
To alleviate this power overestimation for larger filter widths, the disk velocity calculation in Eq.~\eqref{eq:corr_ud} uses a correction factor $M$ derived by \cite{shapiro2019filtered}, which depends on $C_T'$ and the filter width. 
To compute the correction factor $M$, the Taylor series approximation for the ADM correction factor is used \cite[]{shapiro2019filtered} such that 
\begin{equation}
\label{eq:M_corr}
    M = \left( 1 + \frac{C_T'}{4}\frac{1}{\sqrt{3\pi}}\frac{2\Delta}{D}\right)^{-1}. 
\end{equation}

The correction factor given by Eq.~\eqref{eq:M_corr} was derived by \cite{shapiro2019filtered} for yaw aligned actuator disks. 
For low values of $\Delta / D$, the correction factor $M$ has a limited impact on the LES results and the induction and power follow momentum theory \cite[]{shapiro2019filtered}, but low filter widths can also result in numerical oscillations in the flow field due to the ADM forcing discontinuity.
For higher values of $\Delta / D$ with the correction factor implemented for a yaw aligned ADM, the thrust force and power predicted by momentum theory are well reproduced, but the induced velocity in the LES domain does not conform to momentum theory due to the wide force smearing fundamental to the larger values of $\Delta/D$ (see Appendix~\ref{sec:ADM_sensitivity}, Figure~\ref{fig:a_comp_direct}).
In the results presented in \S\ref{sec:results} where analysis of the wake flow field is required, to reduce numerical oscillations in the wake, a larger filter width of $\Delta/D = 3h/(2D)$ is used with the correction factor $M$ given by Eq.~\eqref{eq:M_corr}, where $h = (\Delta x^2 + \Delta y^2 + \Delta z^2)^{1/2}$. 
In the results presented in \S\ref{sec:results} for which only the quantities at the actuator disk are analyzed, a smaller filter width of $\Delta / D = 0.29 h / D = 0.032$ is used such that the correction factor is not required to reproduce the power predicted by momentum theory for the yaw aligned ADM \cite[see also the discussion by][]{shapiro2019filtered}.
In all cases, the ADM thickness is $s = 3\Delta x/2$. 
More discussion of the ADM numerical setup and the interactions between the LES results and the filter width and the correction factor are provided in Appendix~\ref{sec:ADM_sensitivity}.

Simulations are performed with uniform inflow with zero freestream turbulence. 
Periodic boundary conditions are used in the lateral $y$-direction.
A fringe region \cite[]{nordstrom1999fringe} is used in the $x$-direction to force the inflow to the desired profile with a prescribed yaw angle. 
All simulations are performed with a domain $L_x=25D$ in length and cross-sectional size $L_y = 20D, L_z=10D$ with $256\times512\times256$ grid points. 
A large cross-section is used to minimize the influence of blockage on the actuator disk model, which changes as a function of turbine yaw. 
A single turbine is placed inside the domain at the center of the $y$--$z$ plane at a distance $5D$ from the domain inlet in the $x$-direction. 
Simulations are run for two flow-through times $L_x/U_\infty$ to allow the turbine power output to converge, which is sufficient in these zero freestream turbulence inflow cases \cite[]{howland2016wake}.

\section{Results}
\label{sec:results}

In this section, the model predictions are compared to results from LES and the model output is explored to reveal implications for wind farm flow control.
In \S\ref{sec:validation}, the predictive model developed in \S\ref{sec:model} is validated against LES.
The dependence of the induction on the coefficient of thrust is demonstrated in the model and in LES (\S\ref{sec:validation}).
In \S\ref{sec:optimal_ctp}, the model is optimized to find the thrust coefficients which maximize the coefficient of power as a function of the yaw misalignment angle, and the predictions are compared to LES.
Finally, in \S\ref{sec:steering}, the model is used as an initial condition for a turbulent far-wake model. The influence of the induction-yaw relationship developed in \S\ref{sec:model} on a wake steering test case is explored.

\subsection{Comparison between the model and LES}
\label{sec:validation}

\begin{figure}
	\centering
	\begin{tabular}{@{}p{0.33\linewidth}@{\quad}p{0.33\linewidth}@{\quad}p{0.33\linewidth}@{}}
		\subfigimgthree[width=\linewidth,valign=t]{(a)}{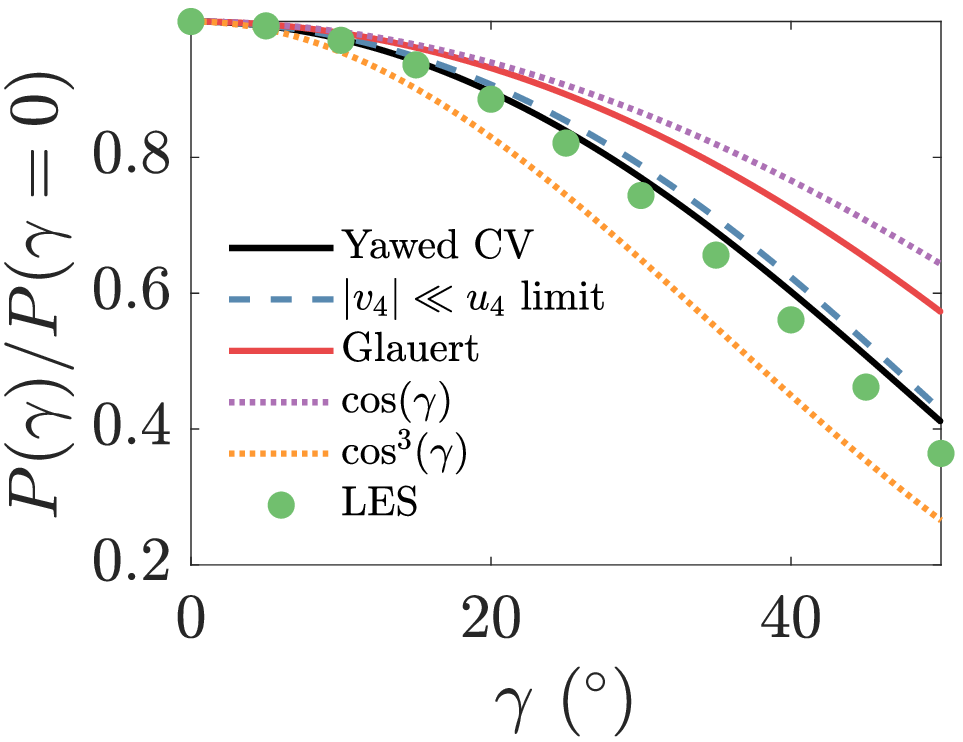} &
		\subfigimgthree[width=\linewidth,valign=t]{(b)}{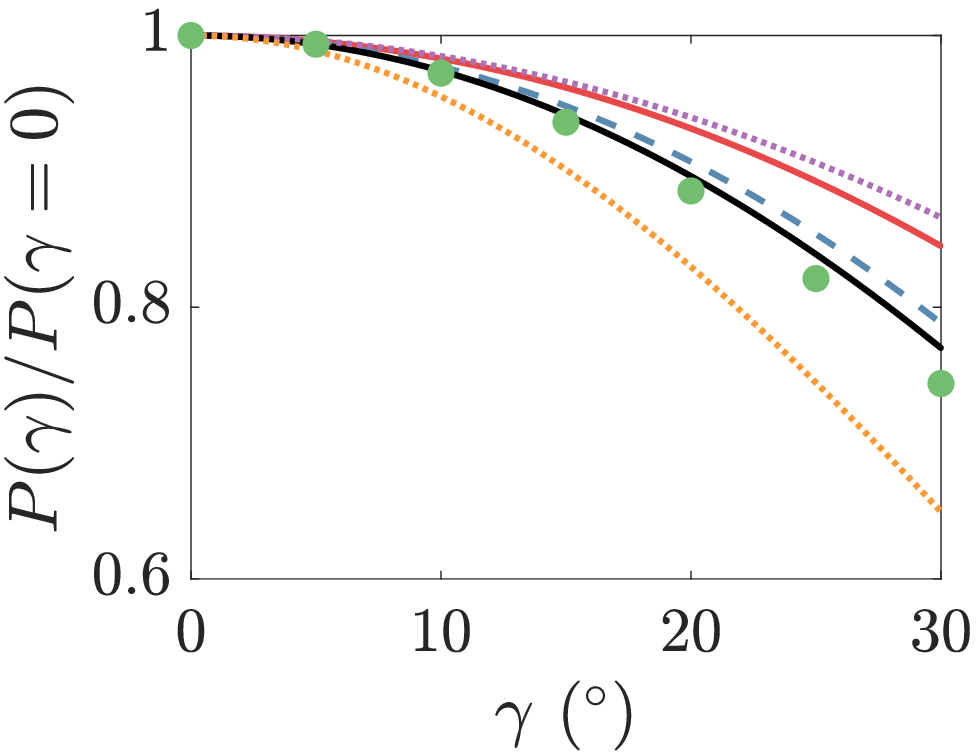} &
		\subfigimgthree[width=\linewidth,valign=t]{(c)}{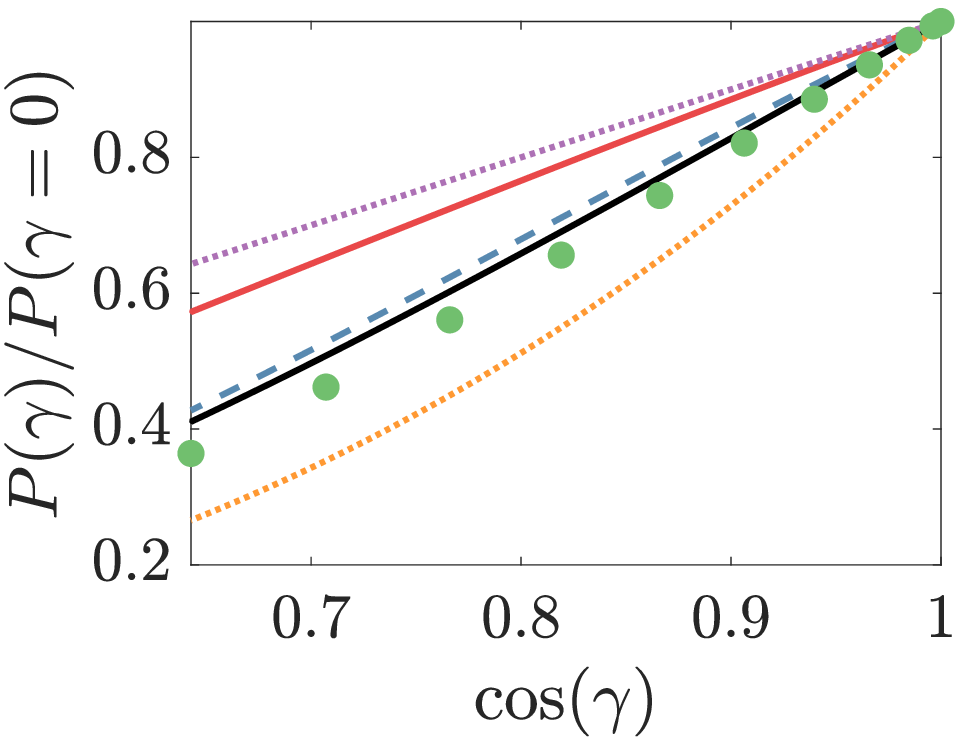}
	\end{tabular}
	\caption{(a) Normalized power production for the yawed actuator disk modeled wind turbine with $C_T^\prime=1.33$, normalized by the power production for a yaw aligned actuator disk modeled wind turbine ($P(\gamma)/P(\gamma=0)$).
	The LES results are shown with green dots.
	The model predictions are given by the \textit{Yawed CV} curve, and the limiting case of $|v_4| \ll u_4$ for the model is shown.
	For reference, $\cos^3(\gamma)$ and $\cos(\gamma)$ curves are shown in addition to the Glauert model (Appendix~\ref{sec:glauert}).
	(b) Zoomed in version of (a) to highlight the performance of different models.
	(c) Same as (a) with $\cos(\gamma)$ on the $x$-axis.
	}
	\label{fig:p_comp}
\end{figure}

The model for $P_r(\gamma)$ (Eq.~\ref{eq:Pr_uniform_analytical_mom_full}) is compared to LES in Figure~\ref{fig:p_comp} for $C_T^\prime=1.33$ for yaw misalignments $0^\circ \leq \gamma \leq 50^\circ$.
The model developed by \cite{glauert1926general}, with the functional form provided in Appendix~\ref{sec:glauert}, and the developed model in the limit of $|v_4| \ll u_4$ (\S\ref{sec:limit_2}) are also shown.
Finally, $\cos(\gamma)$ and $\cos^3(\gamma)$ are shown for reference.
The model developed in \S\ref{sec:cv_full} exhibits the lowest predictive error compared to the LES data.
Neglecting the lateral velocity in the Bernoulli equation, $|v_4| \ll u_4$ (\S\ref{sec:limit_2}), results in a consistent over-prediction of the power production at all yaw misalignment angles because the portion of momentum redistributed to the spanwise velocity, which does not contribute to power, is neglected.
Neglecting the lateral velocity in the Bernoulli equation (Eq.~\eqref{eq:bernoulli}) increases the predicted pressure drop, and therefore the thrust force and the power, because the energy in the spanwise velocity is not accounted for in the outlet flow.

The Glauert model results in a larger power over-prediction. 
This over-prediction is expected, as discussed in \cite{burton2011wind}, since the lift contributions to the thrust in the Glauert model should not contribute to power because it does not contribute to net flow through the disk.
The $\cos(\gamma)$ and $\cos^3(\gamma)$ curves provide upper and lower bounds, respectively, for the LES data and the model predictions.
The commonly assumed $\cos^3(\gamma)$ model \cite[]{burton2011wind} under-predicts the power production because the yaw misalignment reduces the thrust force, which in turn reduces the rotor-normal induction and increases the disk velocity and the power production.

\begin{figure}
	\centering
	\begin{tabular}{@{}p{0.4\linewidth}@{\quad}p{0.4\linewidth}@{}}
	\subfigimgtwo[width=\linewidth,valign=t]{}{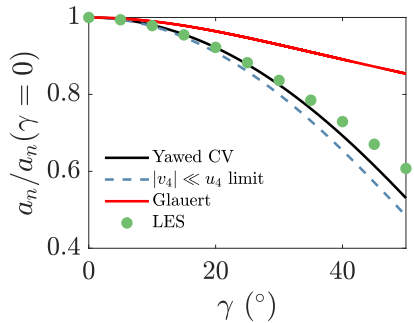}
	\end{tabular}
	\caption{Normalized rotor normal induction for the yawed actuator disk modeled wind turbine with $C_T^\prime=1.33$.
	The model predictions are given by the \textit{Yawed CV} curve, and the limiting case of $|v_4| \ll u_4$ for the model is shown.
	}
	\label{fig:a_comp}
\end{figure}

The model predictions and LES results for the rotor normal induction $a_n(\gamma)$ are shown in Figure~\ref{fig:a_comp}.
As with the power production, the most accurate predictions result from the yawed CV model in \S\ref{sec:cv_full}.
Assuming negligible lateral velocity (\S\ref{sec:limit_2}) results in an under-prediction of the induction, which therefore results in an over-prediction of the disk velocity and the power production (Figure~\ref{fig:p_comp}).
The Glauert model over-predicts the induction, but also over-predicts the power, likely because of the lift contributions to thrust, as mentioned previously.

\begin{figure}
	\centering
	\begin{tabular}{@{}p{0.4\linewidth}@{\quad}p{0.4\linewidth}@{}}
		\subfigimgtwo[width=\linewidth,valign=t]{(a)}{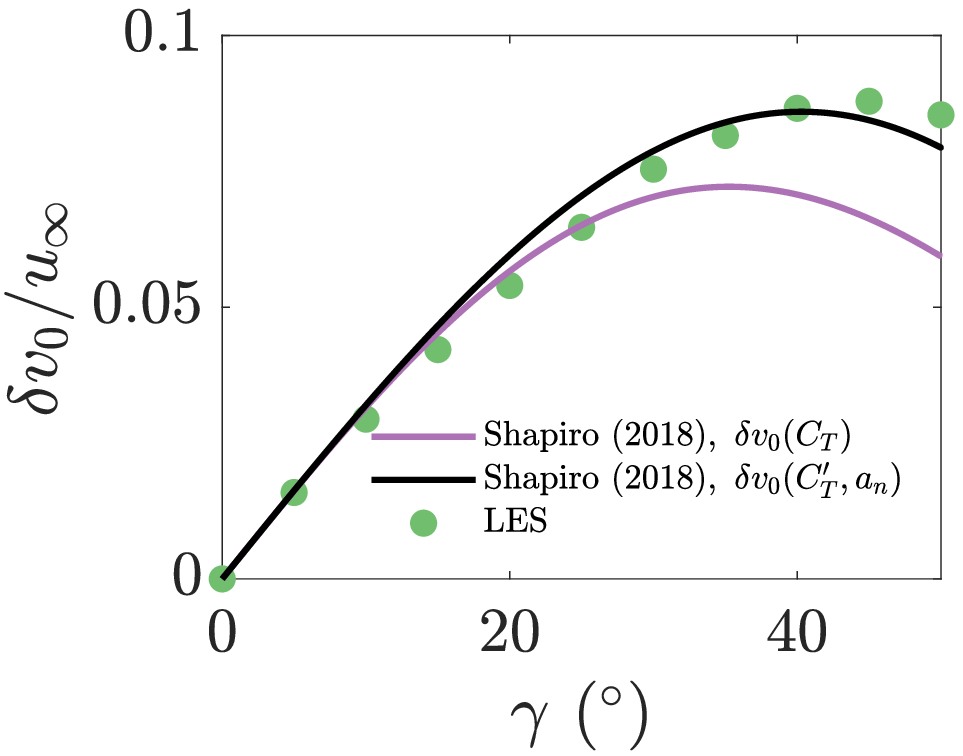} &
		\subfigimgtwo[width=\linewidth,valign=t]{(b)}{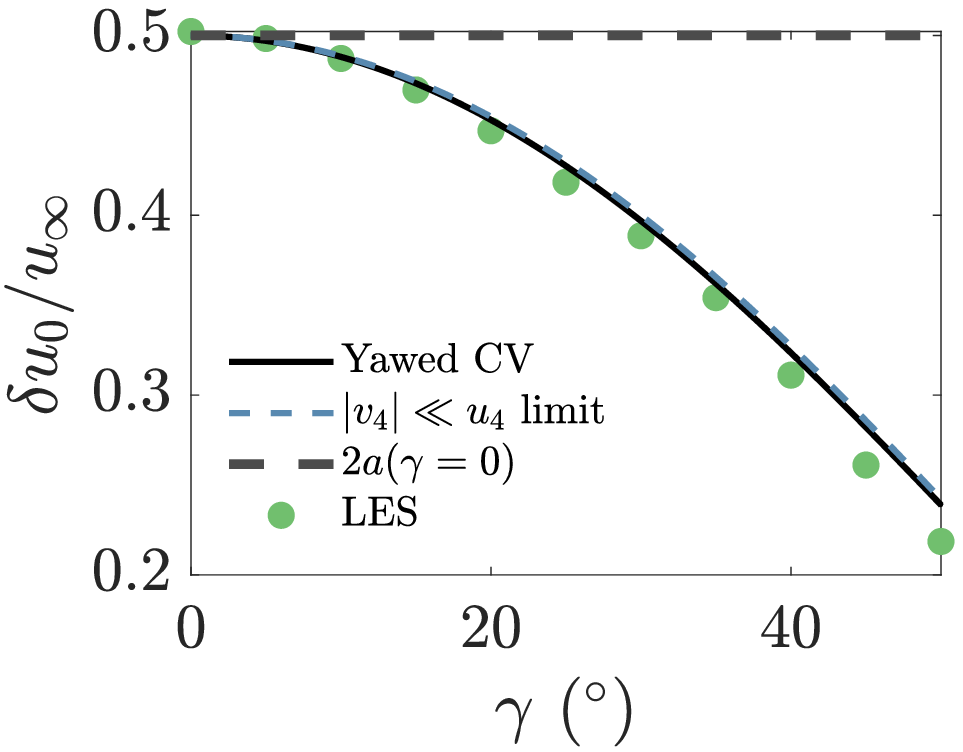}
	\end{tabular}
	\caption{(a) Normalized lateral velocity deficit with $C_T^\prime=1.33$.
	The model predictions for the lateral velocity depending on $C_T$ are shown by $\delta v_0(C_T)$, where $C_T=0.75$.
	The model predictions for the lateral velocity depending on the induction model given by Eq.~\eqref{eq:full_model} and $C_T^\prime$ are shown by $\delta v_0(C_T^\prime, a_n)$.
	(b) Normalized streamwise velocity deficit for the yawed actuator disk modeled wind turbine with $C_T^\prime=1.33$.
	The model predictions are given by the \textit{Yawed CV} curve, and the limiting case of $|v_4| \ll u_4$ for the model is shown.
	}
	\label{fig:u_comp}
\end{figure}

The lateral velocity disturbance, $\delta v_0(\gamma)=v_\infty-v_4=-v_4$, is estimated from LES by averaging the lateral velocity in cross-sections of the actuator disk streamtube \cite[]{shapiro2018modelling}.
The lateral velocity disturbance $\delta v_0(\gamma)$, estimated as the maximum of the cross-sectional averages over $x$, is shown in Figure~\ref{fig:u_comp}(a) along with model predictions.
The maximum value of the lateral velocity disturbance $\delta v_0(\gamma)$ generally occurs approximately $D/2$ downwind of the actuator disk wind turbine.
The original $C_T$-based model of \cite{shapiro2018modelling} ($\delta v_0(\gamma) = \frac{1}{4} C_T u_\infty \cos^2(\gamma) \sin(\gamma)$) under-predicts the initial lateral velocity disturbance at higher yaw angles.
The model developed here yields improved predictions compared to the original model by including the effect of the yaw misalignment on the induction, which the original expression based on $C_T$ does not include. 
Since the induction decreases with increasing magnitude of the yaw angle, the disk velocity will increase.
The increase in disk velocity increases the actuator disk thrust force, partially counteracting the reduction in thrust force from yaw misalignment.
The lateral velocity disturbance based on $C_T^\prime$ and $a_n(\gamma)$ will therefore be larger than a prediction from a model which assumes a fixed $C_T$ as a function of yaw $\gamma$. 

Finally, the streamwise velocity disturbance is shown in Figure~\ref{fig:u_comp}(b).
The LES streamwise velocity disturbance is estimated similarly to $\delta v_0$, although the maximum values of $\delta u_0$ generally occurs approximately $2D$ downwind of the actuator disk.
The streamwise velocity disturbance associated with the yaw aligned wind turbine, $\delta u_0=2a(\gamma=0)$, is shown as a reference.
The streamwise velocity disturbance depends strongly on the yaw misalignment, therefore assuming $\delta u_0(\gamma\neq0)=\delta u_0(\gamma=0)=2a(\gamma=0)$ would yield significant predictive errors in a wake model.
The full model (\S\ref{sec:cv_full}) has slightly improved predictions compared to the limit of negligible lateral velocity (\S\ref{sec:limit_2}), but both model estimates over-predict the streamwise velocity disturbance at larger yaw angles.

\begin{figure}
	\centering
	\begin{tabular}{@{}p{0.33\linewidth}@{\quad}p{0.33\linewidth}@{\quad}p{0.33\linewidth}@{}}
		\subfigimgthree[width=\linewidth,valign=t]{(a)}{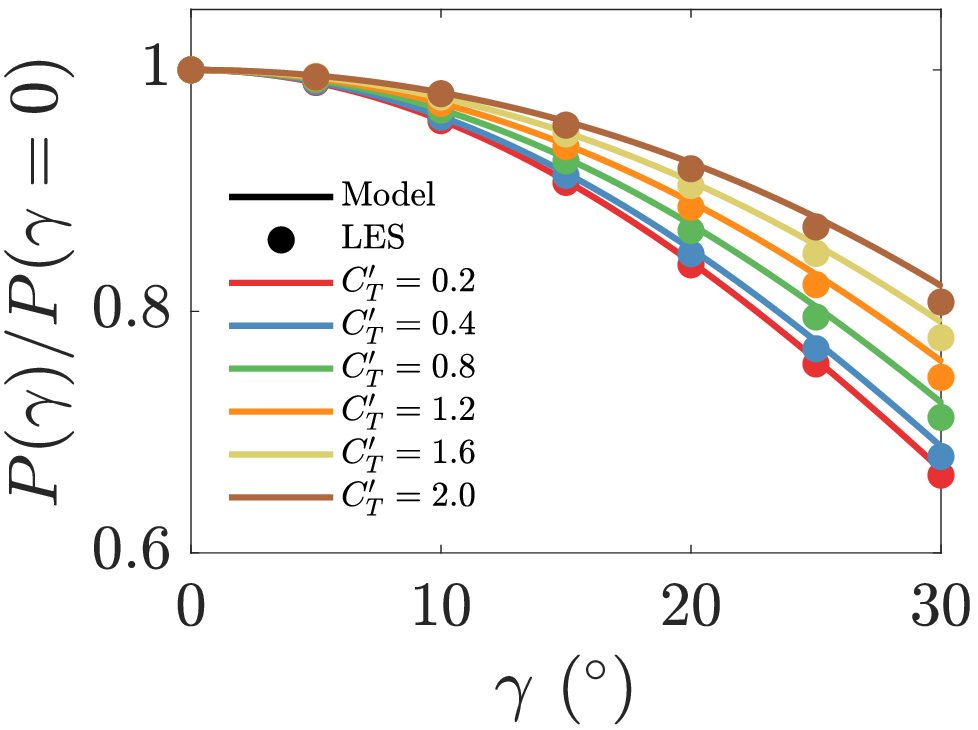} &
		\subfigimgthree[width=\linewidth,valign=t]{(b)}{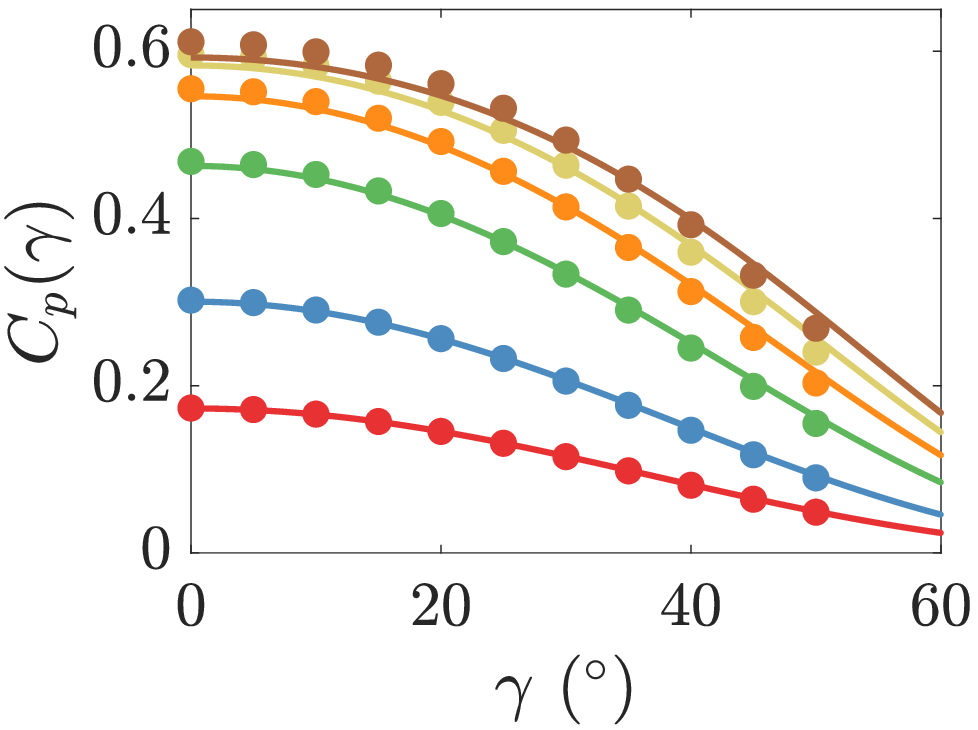} &
		\subfigimgthree[width=\linewidth,valign=t]{(c)}{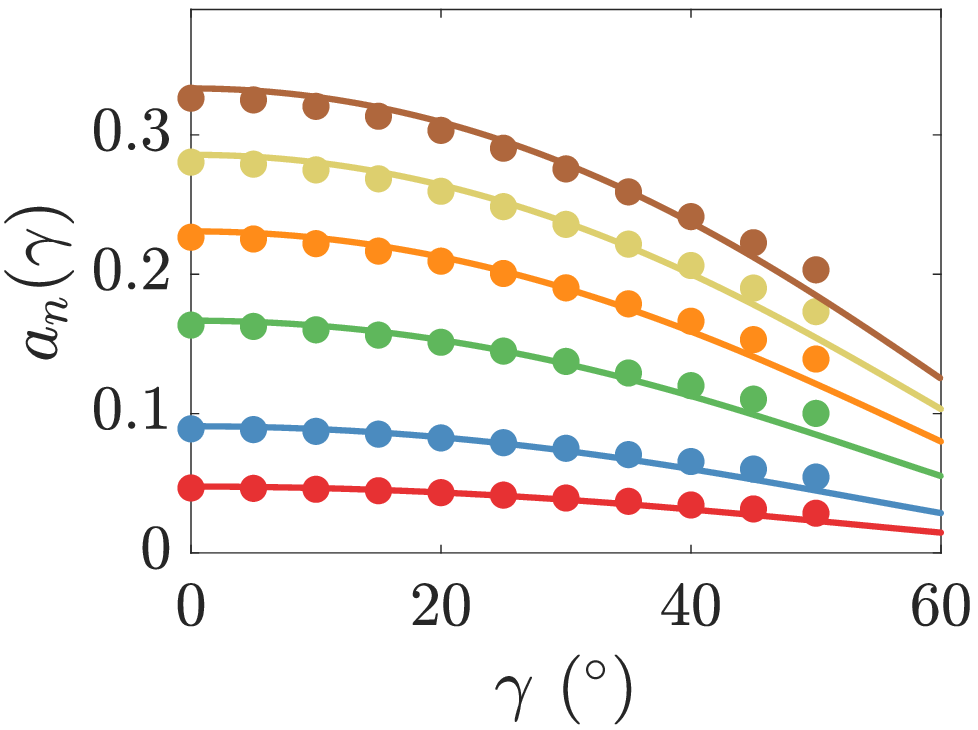} 
	\end{tabular}
	\caption{(a) Power production for the yawed actuator disk modeled wind turbine, normalized by the power production for a yaw aligned actuator disk modeled wind turbine for various values of $C_T^\prime$.
	(b) Coefficient of power $C_p(\gamma)$.
	(c) Rotor-normal induction factor $a_n(\gamma)$.
	}
	\label{fig:p_comp_ctp}
\end{figure}

The model developed in \S\ref{sec:model} reveals that the induction $a_n$, the power $P$, and the power ratio $P_r$ all depend on both the yaw misalignment and the thrust coefficient $C_T^\prime$.
The ADM is simulated in LES over a range of yaw misalignment and $C_T^\prime$ values, where each pair ($\gamma, C_T'$) represents a unique LES case.
The influence of $C_T^\prime$ on the power ratio $P_r$ for the LES data and the model (Eq.~\eqref{eq:Pr_uniform_analytical_mom_full}) is shown in Figure~\ref{fig:p_comp_ctp}(a).
The coefficient of power $C_P$ is shown in Figure~\ref{fig:p_comp_ctp}(b).
The model predictions exhibit low error, compared to the LES data, over a wide range of yaw and thrust values.
We observe that the power reduction by yaw misalignment inherently depends on the value of $C_T^\prime$ (Figure~\ref{fig:p_comp_ctp}(a)), due to the influence of the thrust coefficient $C_T^\prime$ and yaw misalignment on the induction factor $a_n$ (Figure~\ref{fig:p_comp_ctp}(c)).
This result suggests that the power lost due to yaw misalignment in a practical field setting will be turbine-specific, since existing turbine designs operate at a wide range of thrust coefficients \cite[see e.g.][]{hansen2015aerodynamics}.
Further, since the thrust coefficient depends on the operating condition and turbine controller \cite[e.g.][]{ainslie1988calculating}, the power lost due to yaw misalignment will also vary in time for a given turbine design.
Therefore, while an empirically tuned cosine model ($\cos^{P_p}(\gamma)$) may yield a sufficiently small error for a single turbine model and operating condition (e.g. Region II operation \cite[]{pao2009tutorial}), it cannot be expected to extrapolate to other wind turbine designs or control regimes.
Instead, the physics-based model developed in \S\ref{sec:model} can provide a prediction of $P_r(\gamma)$, provided that the thrust force characteristics (i.e. $\vec{F}_T$ or $C_T^\prime$) are known for the turbine model of interest as a function of yaw misalignment.
Future work may integrate the induction-yaw model developed in \S\ref{sec:model} into BEM codes \cite[e.g. FAST,][]{jonkman2005fast}.

\subsection{Optimizing model power and wake deflection in yaw misalignment with $C_T^\prime$}
\label{sec:optimal_ctp}

The induction and power models developed in \S\ref{sec:model} and the results in \S\ref{sec:validation} indicate that the power production of a yaw misaligned actuator disk depends on both the yaw misalignment and the local thrust coefficient $C_T^\prime$. 
In yaw alignment, the well-known Betz limit result estimates that the axial induction factor which maximizes the coefficient of power $C_P = 2 P / (\rho A_d u_\infty^3)$ is $a=1/3$ \cite[e.g.][]{burton2011wind}, with a corresponding value of $C_T^\prime = 2$.
Here, we estimate the value of $C_T^\prime$ which maximizes $C_P$ as a function of yaw misalignment value.
The power produced by the actuator disk is given by Eq.~\eqref{eq:power_ADM}.
The maximum power occurs at $C_T^{\prime*}$ such that $\partial P / \partial C_T^{\prime} = 0$.
Taking the derivative of Eq.~\eqref{eq:power_ADM} with respect to $C_T^\prime$ yields
\begin{equation}
\frac{\partial P}{\partial C_T^\prime} = \frac{1}{2} \rho A_d (1-a_n)^3 \cos^3(\gamma) u_\infty^3 - \frac{3}{2} \rho C_T^\prime A_d (1-a_n)^2 \cos^3(\gamma) u_\infty^3 \frac{\partial a_n}{\partial C_T^\prime}.
\end{equation}
For the full model (Eq.~\eqref{eq:full_model}, \S\ref{sec:cv_full}), $\partial a_n / \partial C_T^\prime$ does not permit a straightforward analytical solution.
To result in an analytical solution, we assume the limit of $|v_4| \ll u_4$ (see \S\ref{sec:limit_2}), giving 
\begin{equation}
\frac{\partial P}{\partial C_T^\prime} = \frac{128 \rho A_d \cos^3(\gamma) u_\infty^3}{(4+C_T^\prime \cos^2(\gamma))^4} \left[1 - \frac{1}{2} C_T^\prime \cos^2(\gamma)\right],
\end{equation}
and power is maximized ($\partial P / \partial C_T^{\prime} = 0$) at
\begin{equation}
C_T^{\prime*}(\gamma) = \frac{2}{\cos^2(\gamma)}.
\label{eq:ctp_max}
\end{equation}
For yaw alignment ($\gamma=0$), the standard Betz limit result is recovered with $C_T^{\prime*}(\gamma=0)=2$.
For yaw misalignment ($\gamma\neq0$), the power maximizing thrust $C_T^{\prime*}(\gamma)$ monotonically increases as a function of increasing yaw misalignment magnitude.
To maximize the power production of a yaw misaligned wind turbine, the turbine should operate at a different thrust coefficient than the standard, optimal Betz value ($C_T=8/9$, $a=1/3$, $C_T^\prime = 2$). 
The maximum power production as a function of the yaw misalignment is
\begin{equation}
P^*(\gamma) = \frac{8}{27} \rho A_d u_\infty^3 \cos(\gamma),
\end{equation}
and the maximum $C_P$ as a function of the yaw misalignment is
\begin{equation}
C_P^*(\gamma) = \frac{16}{27} \cos(\gamma),
\end{equation}
which is equivalent to the Betz limit with an additional factor of $\cos(\gamma)$.
Therefore, subject to the assumptions discussed in \S\ref{sec:model}, the minimum power production lost by a yaw misaligned wind turbine is equal to $\cos(\gamma)$.
As such, $\cos(\gamma)$ represents an upper bound for $P_r(\gamma)$ (Figure~\ref{fig:p_comp}) if $C_T^\prime$ is permitted to change.

\begin{figure}
    \centering
	\begin{tabular}{@{}p{0.33\linewidth}@{\quad}p{0.33\linewidth}@{\quad}p{0.33\linewidth}@{}}
		\subfigimgthree[width=\linewidth,valign=t]{(a)}{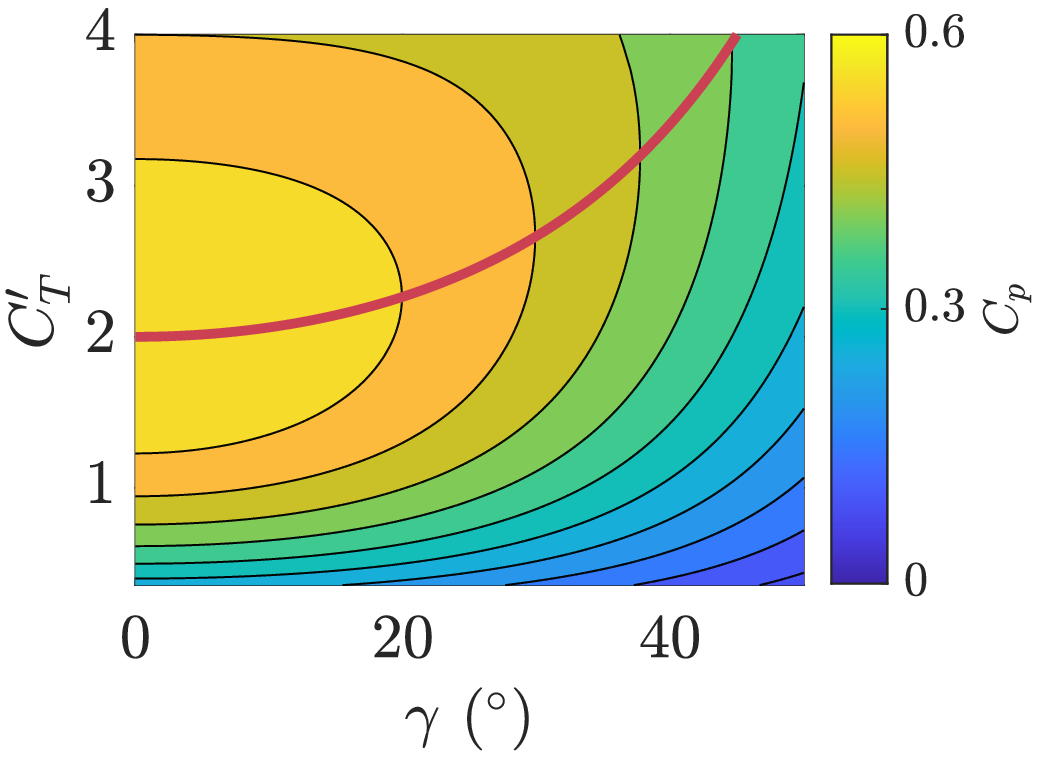} &
		\subfigimgthree[width=\linewidth,valign=t]{(b)}{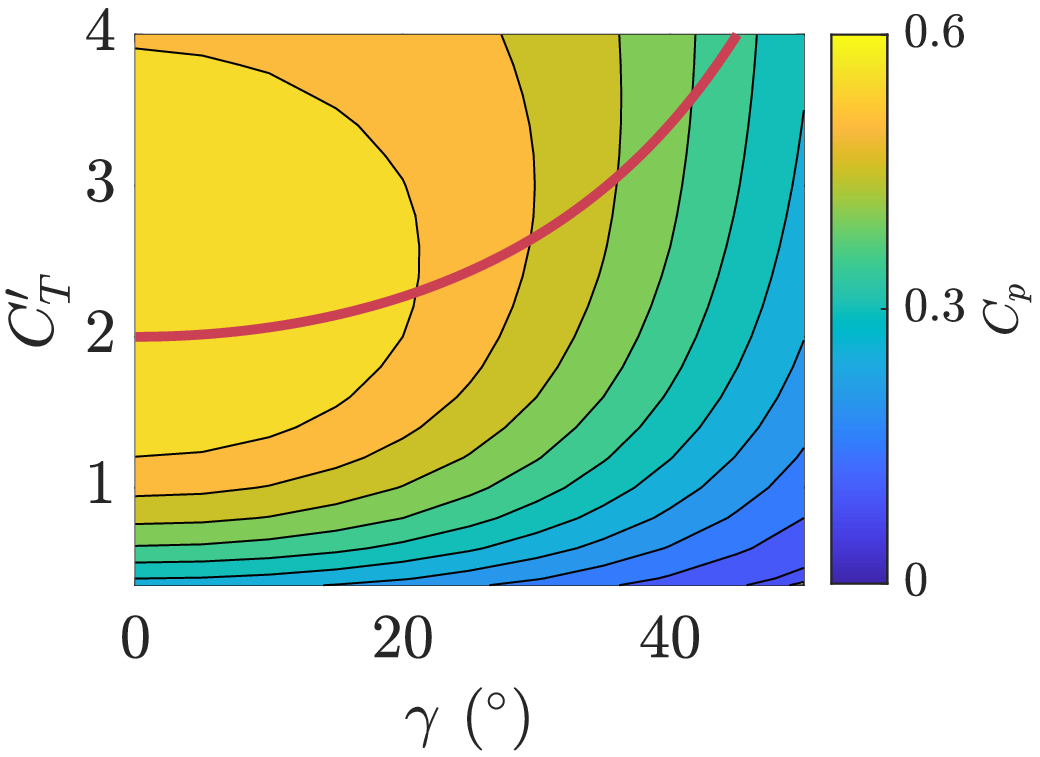} &
		\subfigimgthree[width=\linewidth,valign=t]{(c)}{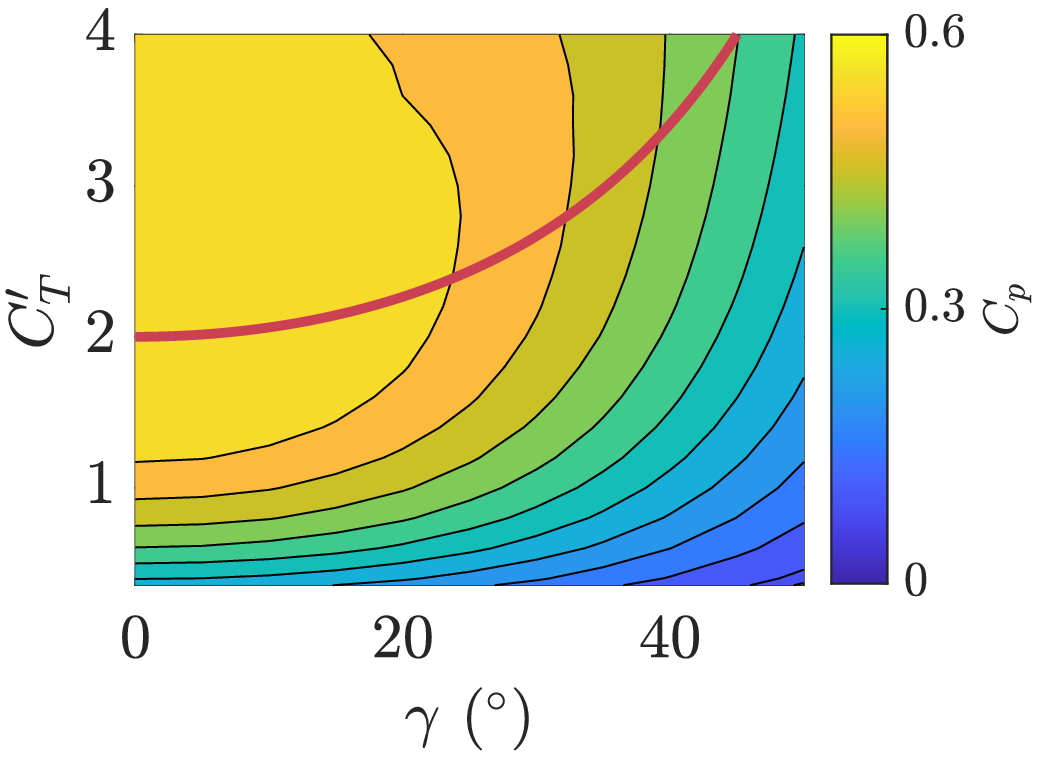}
	\end{tabular}
    \caption{
    (a) Coefficient of power $C_p = 2 P(\gamma) / (\rho A_d u_\infty^3)$ as a function of the yaw misalignment $\gamma$ and thrust coefficient $C_T^\prime$ estimated by the model given in Eqs.~\eqref{eq:full_model} and \eqref{eq:power_ADM}.
    The values of $C_T^\prime$ which maximize power for each yaw misalignment angle are shown by the red line, given as $C_T^{\prime*}(\gamma) = 2/ \cos^2(\gamma)$.
    (b) Same as (a) for LES $C_p$ results with the disk velocity correction factor $M$ given by Eq.~\eqref{eq:M_corr} and $\Delta/D = 3h/(2D)$.
    (c) Same as (b) for LES $C_p$ results with the disk velocity correction factor $M=1$ and $\Delta/D = 0.29h/D=0.032$.
    }
    \label{fig:Cp_max}
\end{figure}

The model predictions (Eq.~\ref{eq:full_model}) for the coefficient of power $C_P$ depending on the yaw misalignment and the thrust coefficient $C_T^\prime$ are shown in Figure~\ref{fig:Cp_max}(a).
Additionally, the optimal thrust coefficient $C_T^{\prime*}(\gamma)$, assuming $|v_4| \ll u_4$, is shown.
The LES coefficient of power $C_P$, for the simulations with the disk velocity $\vec{u}_d$ correction $M$ developed by \cite{shapiro2019filtered} (Eq.~\eqref{eq:M_corr}), is shown in Figure~\ref{fig:Cp_max}(b). 
Figure~\ref{fig:Cp_max}(c) shows the same domain of input yaw misalignment and $C_T^\prime$ for a low value of $\Delta/D$ with $M=1$.
Note that as numerical oscillations in the velocity field worsen with larger shear gradients at the boundary of the wake, the low $\Delta/D$ LES contours in Figure~\ref{fig:Cp_max}(c) become less smooth (and accurate) as $C_T'$, and therefore $\delta u_0$, increases. 
There are similar qualitative trends in the LES $C_P$ compared to the model predictions in Figure~\ref{fig:Cp_max}(a), especially for $C_T^\prime \lesssim 2$.
As demonstrated in Figure~\ref{fig:p_comp_ctp}, the model predicts the LES output quantitatively well.
The differences between the model predictions and LES values of $C_P$ generally increase with increasing $C_T^\prime$.
One cause of discrepancy, in addition to potential modeling simplifications in \S\ref{sec:model}, is that the ADM implementation in LES is known to underestimate wind turbine induction \cite[]{munters2017optimal,shapiro2019filtered}. 
Consequently, the maximum coefficient of power in LES is $C_P = 0.602$, even with the correction factor used, which is higher than the Betz limit ($0.593$).

Following a similar procedure, the thrust coefficient value which maximizes the magnitude initial lateral velocity $|v_4|$, and therefore the wake deflection, is $C_T^{\prime}(\gamma) = 4/\cos^2(\gamma)$.
However, these values of $C_T^{\prime}(\gamma)$ produce inductions which are greater than one, which is inconsistent with the momentum theory based model in Eq.~\eqref{eq:full_model}.
Therefore, for realizable values of $C_T^\prime$, the lateral velocity magnitude $|v_4|$ is a monotonically increasing function of $C_T^\prime$.
Conversely, $u_4$, the streamwise wake velocity, is a monotonically decreasing function of $C_T^\prime$.

\begin{figure}
    \centering
	\begin{tabular}{@{}p{0.45\linewidth}@{\quad}p{0.45\linewidth}@{}}
		\subfigimgthree[width=\linewidth,valign=t]{(a)}{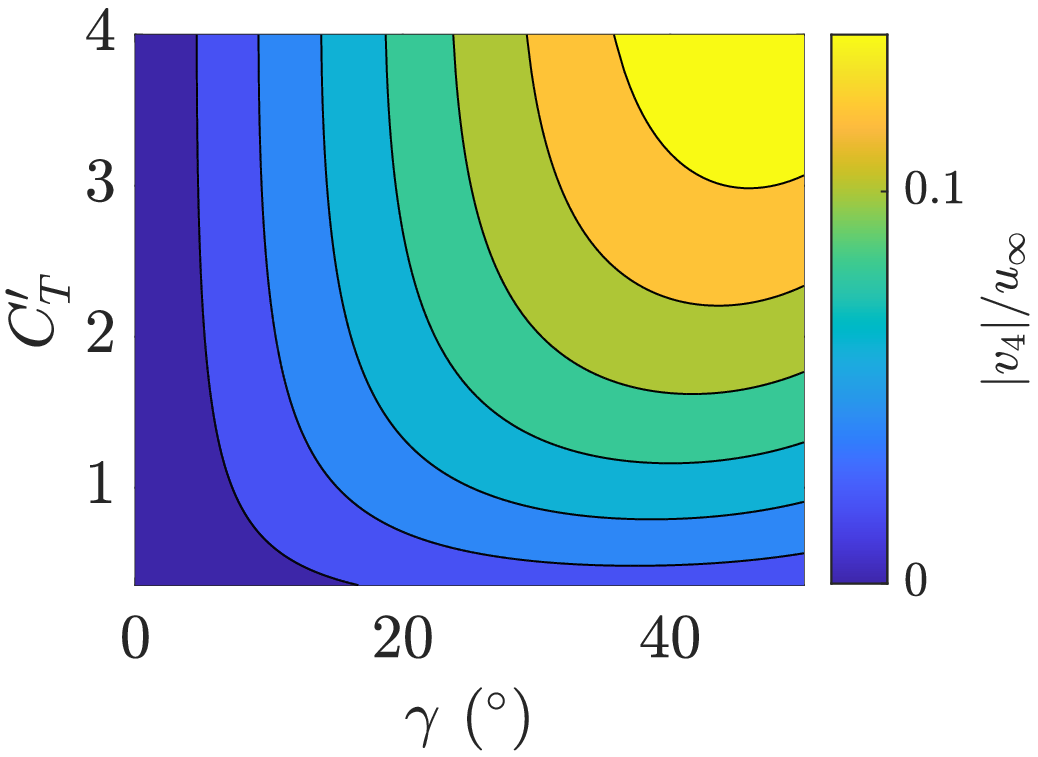} &
		\subfigimgthree[width=\linewidth,valign=t]{(b)}{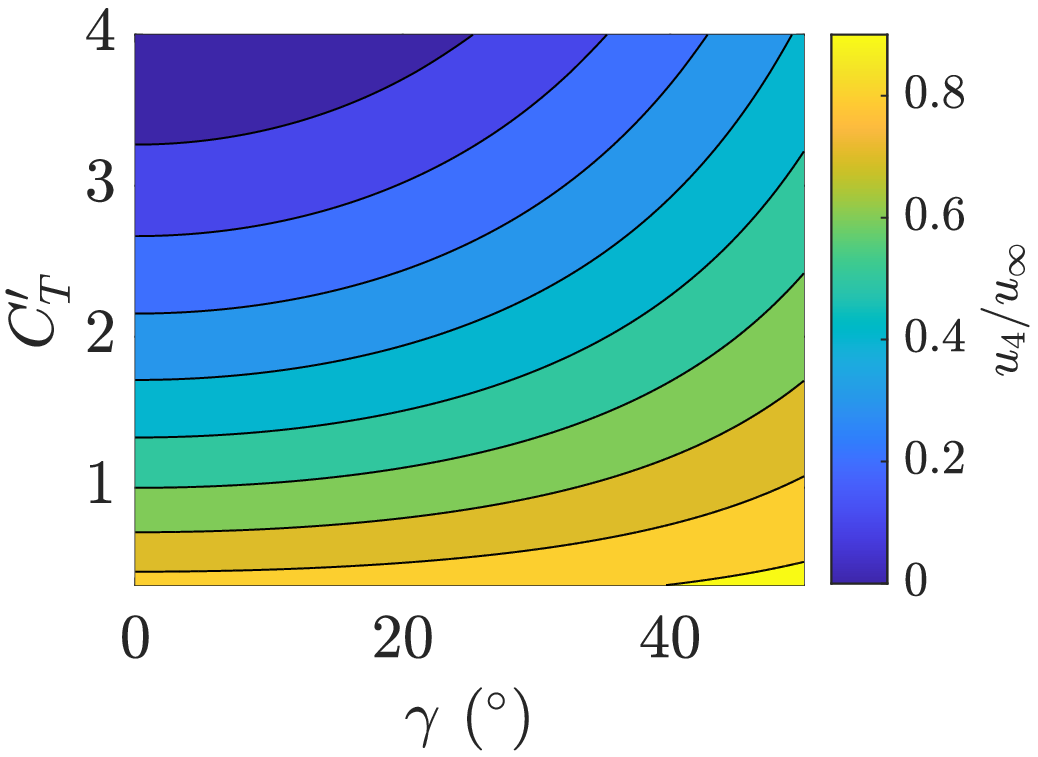}
	\end{tabular}
    \caption{
    (a) Magnitude of the initial lateral velocity $|v_4|/u_\infty$ as a function of the yaw misalignment $\gamma$ and thrust coefficient $C_T^\prime$ estimated by the model given in Eq.~\eqref{eq:full_model}.
    (b) Initial streamwise velocity $u_4/u_\infty$ as a function of the yaw misalignment $\gamma$ and thrust coefficient $C_T^\prime$ estimated by the model given in Eq.~\eqref{eq:full_model}.
    }
    \label{fig:u4_v4_model}
\end{figure}

The model-predicted normalized streamwise and spanwise outlet velocities are shown in Figures~\ref{fig:u4_v4_model}(a) and \ref{fig:u4_v4_model}(b), respectively.
While the power production reveals a non-monotonic trend and permits an optimal set of thrust coefficients ($C_T^{\prime*}(\gamma)$), both $u_4$ and $|v_4|$ show monotonic behavior for realizable values of $C_T^\prime$.
For wake steering, the power production of a waked turbine will depend on both the streamwise wake velocity ($u_4$), and the wake deflection (integrated form of $v_4$).
Notably, the wake deflection is an increasing function of $C_T^\prime$ (Figure~\ref{fig:u4_v4_model}(a)), but the velocity deficit is also an increasing function of $C_T^\prime$ (Figure~\ref{fig:u4_v4_model}(b)).
Therefore, the value of $C_T^\prime$ which maximizes the power production of a waked downwind turbine will depend on the wind farm and flow configuration.
In \S\ref{sec:steering}, we explore this dependency in an analytical, turbulent wake model which uses the inviscid model developed in \S\ref{sec:model} as an initial condition.

\subsection{Implications for wake steering and induction control}
\label{sec:steering}

The impact of the yaw misalignment $\gamma$ on the rotor normal induction $a_n$ will impact the power production, wake deflection, and wake velocity deficit of a yaw misaligned turbine.
All three of these effects will modify the performance of wake steering control (intentional yaw misalignment).
Similarly, as demonstrated in \S\ref{sec:validation} and \S\ref{sec:optimal_ctp}, changing the local thrust coefficient $C_T^\prime$ (often called induction control) will also influence the power and wake properties of a yaw misaligned turbine.
In this section, we assess the role of yaw and thrust modifications on combined wake steering and induction flow control.

To assess the role of the developed induction model on wake steering and induction-based wind farm flow control, the model (see \S\ref{sec:model}, Eq.~\eqref{eq:full_model}) is used as an initial condition for a turbulent far-wake model.
Inviscid near-wake models are commonly used as initial conditions for far-wake models \cite[e.g.][]{frandsen2006analytical, bastankhah2016experimental, shapiro2018modelling}.
A Gaussian far-wake model is used, and the full model form is provided in Appendix~\ref{sec:far}.

We consider a simplified wind turbine array with two wind turbines spaced with streamwise and spanwise separation of $S_x=8D$ and $S_y=0.5D$, respectively.
Given the spanwise spacing of $S_y=0.5D$, positive yaw misalignments (counter-clockwise rotation viewed from above) will be preferable to negative yaw \cite[e.g.][]{howland2022collective}.
For illustrative purposes, the wake model parameters, the wake spreading rate and the proportionality constant of the presumed Gaussian wake, are set to representative values in the literature of $k_w=0.07$ \cite[]{stevens2015coupled, howland2020optimal} and $\sigma_0=0.25$ \cite[]{shapiro2019wake}, respectively.
We vary the yaw misalignment $\gamma_1$ and the thrust coefficient $C_{T,1}^\prime$ of the leading freestream turbine.
The yaw misalignment and thrust coefficient for the downwind turbine are held constant at the individual power maximization levels of $\gamma_2=0^\circ$ and $C_{T,2}^\prime=2$, respectively.

We consider the wind farm efficiency as a function of the yaw misalignment and the thrust coefficient of the leading turbine.
The wind turbine efficiency $\eta_i$ for turbine $i$ is given by
\begin{equation}
\eta_i(\gamma_1, C_{T,1}^\prime) = \frac{P_i(\gamma_1, C_{T,1}^\prime)}{ \frac{1}{2} \rho A_d u_\infty^3}.
\label{eq:efficiency}
\end{equation}
Equation~\eqref{eq:efficiency}, which is a nondimensional representation of the power production, differs from $C_p$ because it is based on the freestream wind speed $u_\infty$ for both freestream and waked turbines.
The power production of each turbine is estimated as
\begin{equation}
P_i(\gamma) = \frac{1}{2} \rho C_{T,i}^\prime A_d \left[(1-a_{n,i}(\gamma_i,C_{T,i}^\prime)) \cos(\gamma_i) u_{e,i} \right]^3,
\end{equation}
where $u_e$ is the rotor-averaged velocity accounting for wake interactions (more details provided in Appendix~\ref{sec:far}).
The wind farm efficiency is $\eta = \sum_{i=1}^{N_t} \eta_i / N_t$, where $N_t$ is the number of wind turbines.

\begin{figure}
	\centering
	\begin{tabular}{@{}p{0.33\linewidth}@{\quad}p{0.33\linewidth}@{\quad}p{0.33\linewidth}@{}}
		\subfigimgtwo[width=\linewidth,valign=t]{(a)}{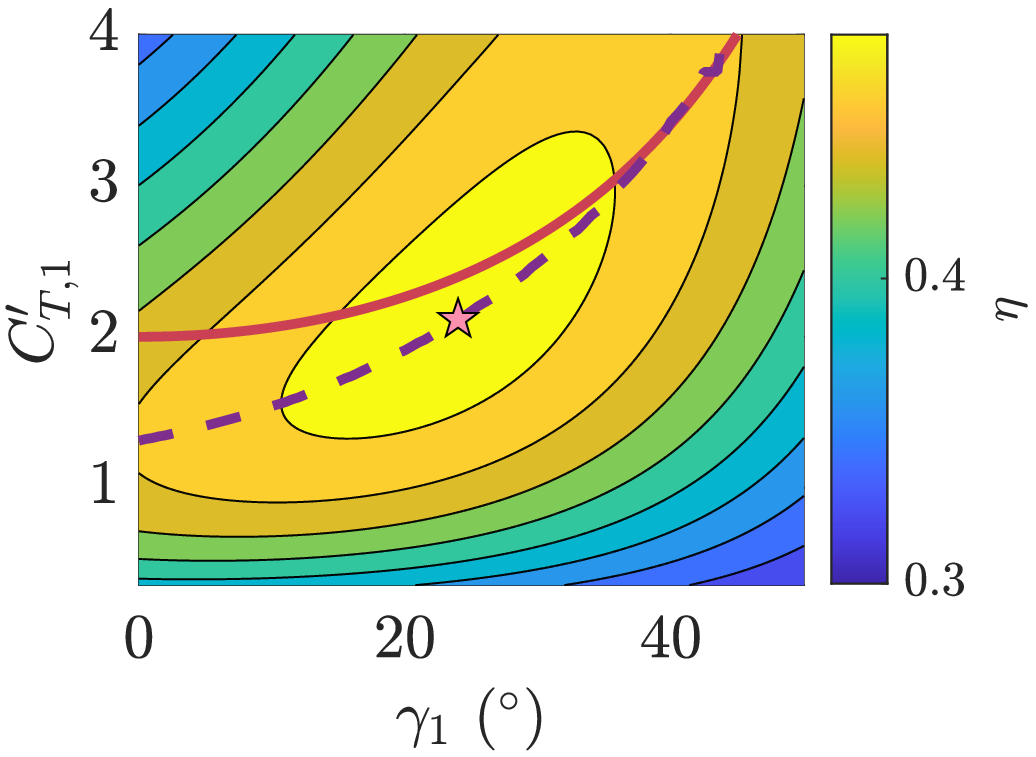} &
		\subfigimgtwo[width=\linewidth,valign=t]{(b)}{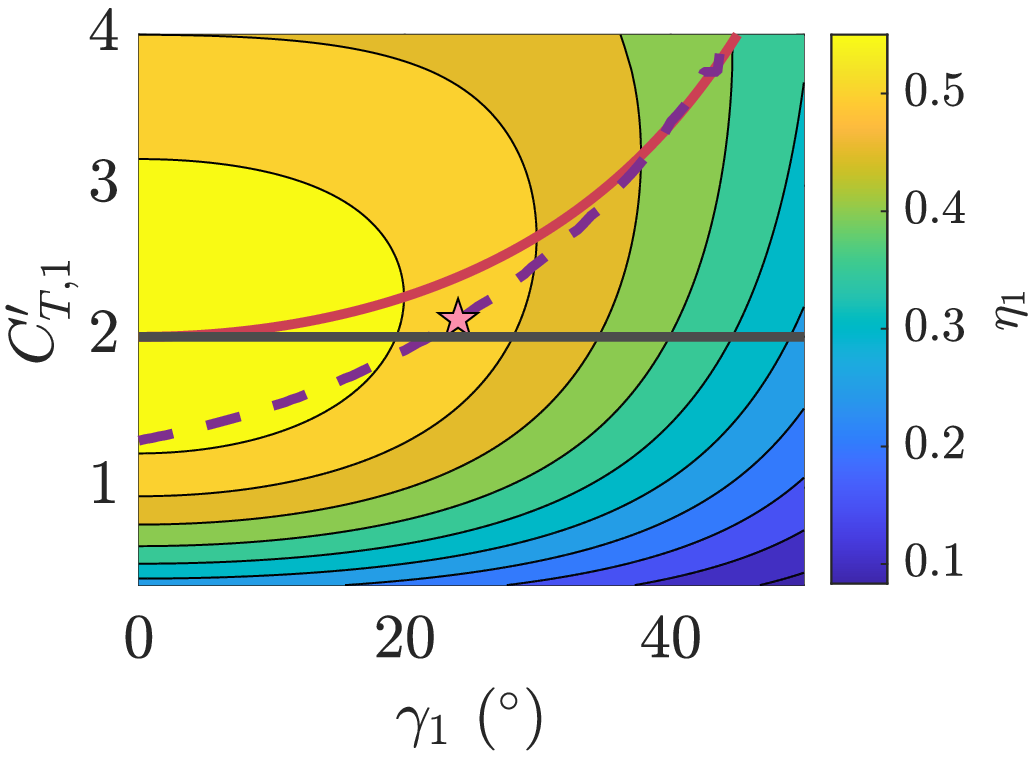} &
		\subfigimgtwo[width=\linewidth,valign=t]{(c)}{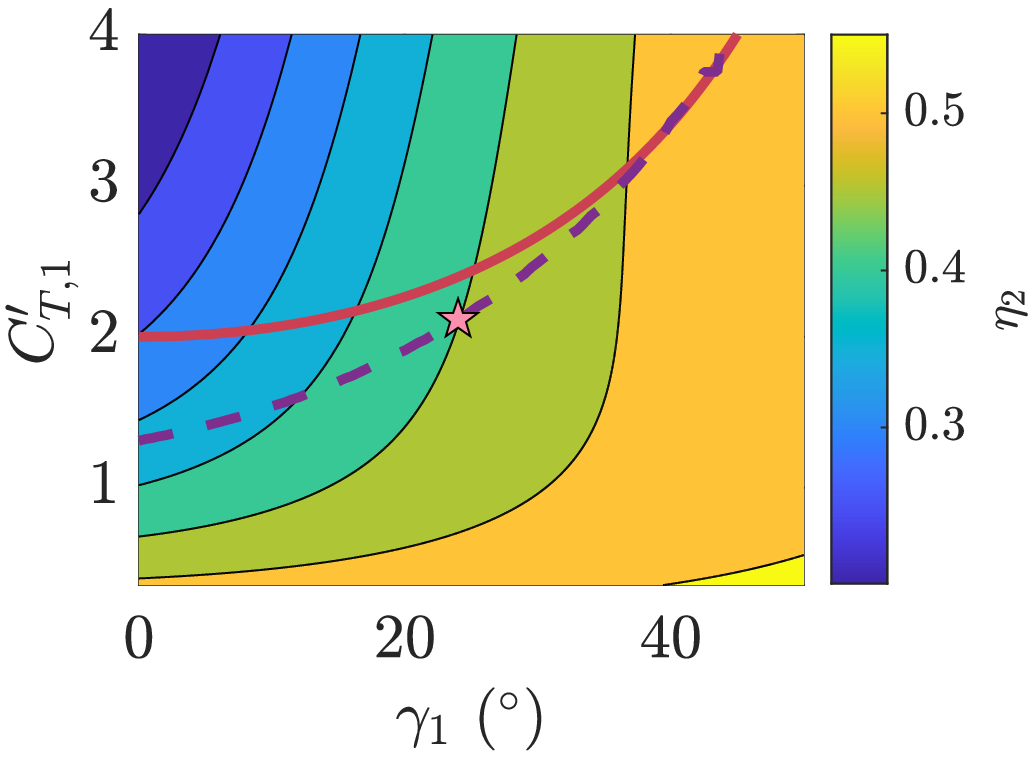}
	\end{tabular}
	\caption{
	Wake model predictions for the (a) total wind farm efficiency $\eta$, (b) freestream individual turbine $1$ efficiency, and (c) waked individual turbine $2$ efficiency.
	The efficiency is calculated using Eq.~\eqref{eq:efficiency}.
	The location of maximum array efficiency (highest total array power output) is indicated with a star symbol.
	The values of $C_T^\prime$ which maximize the total array power (turbines $1$ and $2$) for each yaw misalignment angle are shown by the dashed purple line, as predicted empirically from the wake model output.
	The values of $C_T^\prime$ which maximize the freestream turbine power (turbine $1$) for each yaw misalignment angle are shown by the red line, and given as $C_T^{\prime*}(\gamma) = 2/ \cos^2(\gamma)$.
	In (b), the individual freestream yaw-aligned wind turbine efficiency maximizing thrust coefficient $C_T^{\prime*}(\gamma=0)=2$ is shown with a horizontal line.
	}
	\label{fig:steering_eta}
\end{figure}

The total wind turbine array efficiency $\eta$ is shown as a function of $\gamma_1$ and $C_{T,1}^\prime$ in Figure~\ref{fig:steering_eta}(a), with the array efficiency maximizing point denoted with the star symbol.
We can make a few observations.
First, we note that the maximum array efficiency does not occur at $\gamma_1=0^\circ$ and $C_T^\prime=2$, the optimal settings for an individual turbine, meaning that the array efficiency can be increased through flow control.
The array efficiency maximizing value of $C_{T,1}^\prime$ at each yaw misalignment value is shown by a dashed line in Figure~\ref{fig:steering_eta}(a).

Second, the maximum array efficiency is also not located directly on the individual turbine $C_P$ maximizing curve (see \S\ref{sec:optimal_ctp}) of $C_{T,1}^{\prime*}(\gamma)=2/\cos^2(\gamma)$.
In particular, the array efficiency maximizing values of $C_{T,1}^\prime$ are always below the turbine $1$ efficiency maximizing values given by $C_{T,1}^{\prime*}(\gamma)=2/\cos^2(\gamma)$.
At low yaw misalignment values, the array and individual turbine maximizing values of $C_{T,1}^\prime$ differ the most and this difference decreases with increasing yaw misalignment angles.
The efficiency of turbine $1$ is shown in Figure~\ref{fig:steering_eta}(b).
As is shown, the array efficiency is maximized at a turbine $1$ yaw and thrust which is neither the standard Betz maximum nor the maximum as a function of $\gamma$ (Eq.~\eqref{eq:ctp_max}).
While operating turbine $1$ at $C_{T,1}^{\prime*}(\gamma)$ would maximize the turbine $1$ power, given the applied yaw misalignment, this operation also results in larger wake velocity deficits.
The efficiency of turbine $2$ depending on $\gamma_1$ and $C_{T,1}^\prime$ is shown in Figure~\ref{fig:steering_eta}(c).
The efficiency of turbine $2$ increases with increasing turbine $1$ yaw or a decreasing turbine $1$ thrust coefficient.
In summary, the array efficiency maximizing operation has a lower value of $C_{T,1}^\prime$ than the value which maximizes the power of turbine $1$, in order to increase the power of turbine $2$.
The operating point of optimal efficiency is a combination of yaw and induction control.
At lower yaw values, induction control (reduction in $C_{T,1}^\prime$) is more active.
At higher yaw values, the array efficiency is maximized at values of $C_{T,1}^\prime$ which are close to the operation which maximizes the upstream turbine efficiency ($C_{T,1}^{\prime*}(\gamma)=2/\cos^2(\gamma)$).

The proximity of the optimal operating point to the turbine $1$ power maximizing curve ($C_T^{\prime*}(\gamma)=2/\cos^2(\gamma)$) reaffirms that wake steering control is strongly dependent on the power-yaw relationship of the freestream turbine, since the freestream turbines contribute a larger fraction of the total array power \cite[]{howland2020influence, howland2022collective}.
However, the departure from the curve (i.e. the misalignment of the solid red line and the dashed purple line in Figure~\ref{fig:steering_eta}(a)) reaffirms that it is also important to accurately model the wakes and the power of each turbine in the array to locate the array power-maximizing operation.
The power-maximizing operation will depend on the wind conditions and the wind farm geometry, necessitating an accurate parametric model which can capture these trends.
The model developed in \S\ref{sec:model} and used here enables the prediction of the induction, thrust, and power of a yaw misaligned actuator disk, in addition to the velocity deficit initial conditions for far-wake models.

\section{Conclusions}
\label{sec:conclusions}

The velocity induced by an actuator disk depends jointly on the yaw misalignment angle and the thrust coefficient.
This dependence affects the thrust, wake velocity deficit, wake deflection, and power production of a yaw misaligned actuator disk.
Therefore, the characteristic reduction in power production associated with wind turbine yaw misalignment depends on the thrust coefficient of the wind turbine.
As such, a tuned, empirical cosine model ($P_r = \cos^{P_p}(\gamma)$) for the power-yaw relationship of a wind turbine is inherently turbine model-specific.
Specifically, the empirical power-yaw factor $P_p$ can only be potentially reasonable for turbines with the same thrust coefficient, although we note that the relative error of a cosine-based model increases with increasing yaw angles since the true form of $P_r$ is not exactly a cosine function.

An analytical model for the induction of a yaw misaligned actuator disk is developed and validated against large eddy simulations of an actuator disk model wind turbine.
The model yields improved quantitative predictions of the induction, velocities, and power of a yawed actuator disk, compared to existing models, by accounting for the effect of the induction on the wind turbine thrust and the momentum associated with the lateral velocity at the outlet of the streamtube encompassing the disk.
We optimize the coefficient of power predicted by the developed model to find the thrust coefficient which maximizes the power production of a yawed actuator disk for each value of the yaw misalignment angle.
The optimization results, which are the yawed actuator disk analogue to the classical Betz limit, demonstrate that the thrust coefficient should increase monotonically with an increasing magnitude of yaw misalignment to track the optimal power production ($C_T^{\prime*}(\gamma) = 2/\cos^2(\gamma)$) and that the maximum power produced by an individual yaw misaligned actuator disk is $C_P^*(\gamma) = \frac{16}{27} \cos(\gamma)$.

Finally, the developed induction model is used as an initial condition for a turbulent far-wake model to explore an example, two-turbine wind farm control scenario.
The model-predicted combined power production for the two turbine array is maximized through a combination of yaw (wake steering) and thrust coefficient (induction) control modifications which deviate from the individual turbine power-maximizing operation ($C_T^\prime=2$, $\gamma=0^\circ$).
The yaw and thrust coefficient of the leading turbine affect its own power production (power-yaw relationship $P_r(\gamma)$) but also affect the wake velocity deficit and wake deflection, which influences the power production of the downwind turbine.
The modeling results demonstrate the physical mechanisms for synergistic wake steering and induction, a strategy which has been shown to be effective in previous simulation studies of farm flow control \cite[e.g.][]{munters2018dynamic}.

For rotational, utility-scale wind turbines, the realized power-yaw relationship (i.e. $P_r(\gamma)$) will depend on the realized local thrust coefficient $C_T^\prime$ and any potential dependence of $C_T^\prime$ on the yaw misalignment angle.
Such a dependence can be integrated into the present modeling framework through the functional form of the thrust force (Eq.~\eqref{eq:ft_full_les}).
In addition to the effects of the yaw and thrust coefficient on the rotor-averaged induction, yaw misalignment also generates an induced velocity which exhibits spatial variation over the rotor area \cite[]{hur2019review}.  
Future work that focuses on extending the present analysis to rotational wind turbines should consider the effects of spatially variable induction.

Often, numerical implementations of blade-element momentum (BEM) theory predict that the power ratio of a yaw misaligned wind turbine follows $P_r = \cos^3(\gamma)$ \cite[e.g.][]{liew2020analytical}.
Yawed wind turbines, operating with a fixed $C_T^\prime$ in uniform flow, will not have a power ratio of $P_r = \cos^3(\gamma)$ since the rotor-normal induction factor is reduced by the yaw misalignment. 
The power produced by a yaw misaligned turbine is therefore greater than $P_r = \cos^3(\gamma)$ (i.e. $P_p<3$), although the particular value of power lost by yaw will depend on $C_T^\prime$.
Future work should incorporate the induction model developed here into BEM solvers.
Finally, this study focused on spatially uniform inflow.
Wind speed and direction shear \cite[]{howland2020influence} and wake interactions \cite[]{liew2020analytical} affect the power production of yaw misaligned wind turbines.
Future work should consider the effects of wind speed and direction shear on the induced velocity of a yawed actuator disk.

\section*{Acknowledgements}
K.S.H. and M.F.H. acknowledge funding from the National Science Foundation (Fluid Dynamics program, grant number FD-2226053).
H.M.J. acknowledges support from Siemens Gamesa Renewable Energy.
M.F.H. gratefully acknowledges partial support from the MIT Energy Initiative and MIT Civil and Environmental Engineering.
All simulations were performed on Stampede2 supercomputer under the XSEDE project ATM170028.
The authors thank Aditya Aiyer and Carl Shapiro for insightful discussions during the beginning of this study.

\section*{Declaration of Interests}
The authors report no conflict of interest.
 
\FloatBarrier

\appendix

\section{Sensitivity of LES ADM induction to numerical setup}
\label{sec:ADM_sensitivity}

\begin{figure}
	\centering
	\begin{tabular}{@{}p{0.4\linewidth}@{\quad}p{0.4\linewidth}@{}}
	\subfigimgtwo[width=\linewidth,valign=t]{(a)}{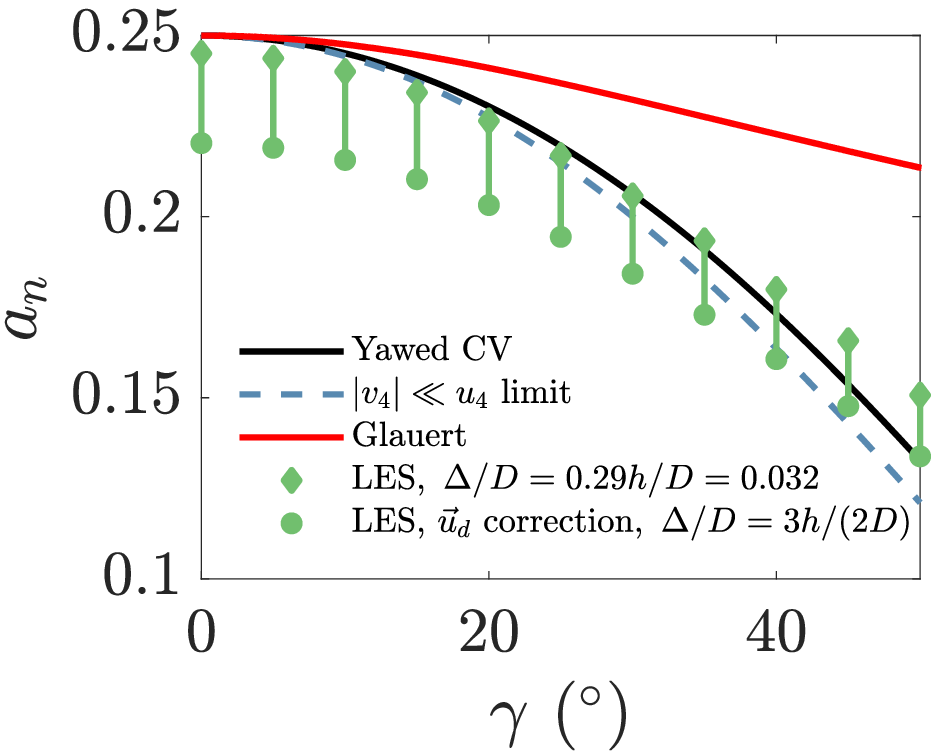} & 
	\subfigimgtwo[width=\linewidth,valign=t]{(b)}{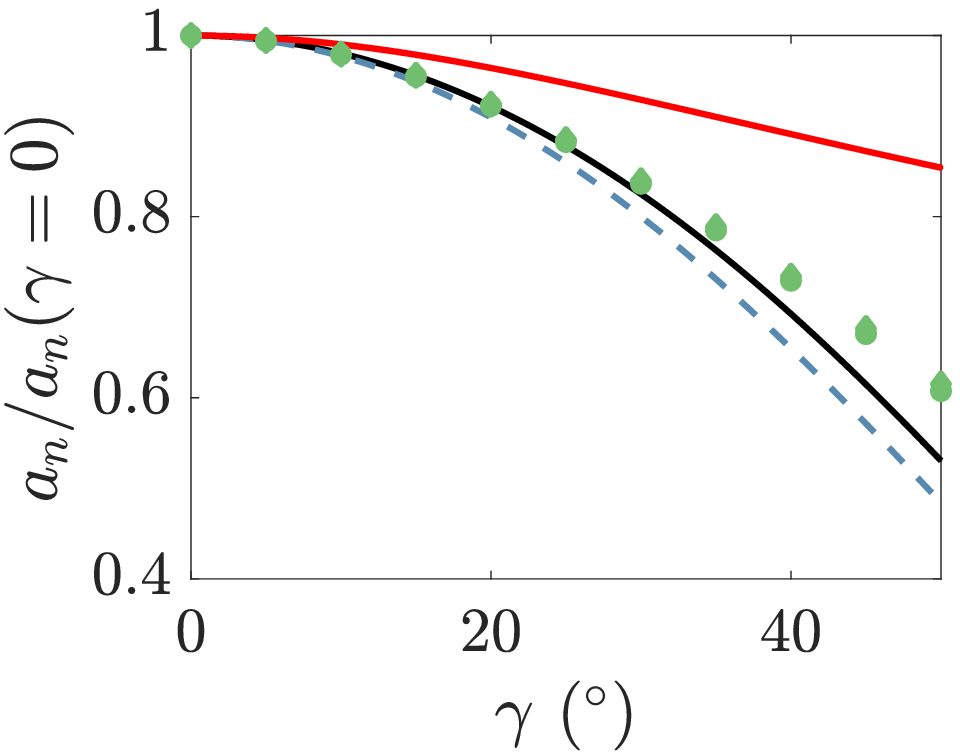}
	\end{tabular}
	\caption{(a) Rotor-normal induction for the yawed actuator disk modeled wind turbine with $C_T^\prime=1.33$.
	The LES results are shown with ($M$ given by Eq.~\eqref{eq:M_corr}, $\Delta/D=3h/(2D)$) and without ($M=1$, $\Delta/D=0.29h/D=0.032$) the disk velocity $\vec{u}_d$ correction factor. 
	The cases with and without the correction factor use a larger and smaller filter width $\Delta$, respectively.
	The model predictions are given by the \textit{Yawed CV} curve, and the limiting case of $|v_4| \ll u_4$ is shown.
	(b) Same as (a) except the rotor-normal induction values are normalized by the yaw aligned ADM rotor-normal induction, $a_n(\gamma)/a_n(\gamma=0)$.
	}
	\label{fig:a_comp_direct}
\end{figure}

The rotor-normal induction factor $a_n$ for $C_T^\prime=1.33$ is shown as a function of the yaw misalignment angle $\gamma$ in Figure~\ref{fig:a_comp_direct}(a) for LES cases with ($M$ given by Eq.~\eqref{eq:M_corr}) and without ($M=1$) the disk velocity correction factor $M$.
For the yaw aligned ADM, the uncorrected disk velocity simulation with a small filter width $\Delta/D=0.29h/D=0.032$ approaches the momentum theory estimate of $a_n=0.25$, ($\hat{a}_n(\gamma=0)=0.245$).
On the other hand, the larger filter width case, $\Delta/D=3h/(2D)$, results in an under-prediction of the momentum theory induction ($\hat{a}_n(\gamma=0)=0.220$), even with the disk velocity correction activated.
While the smaller filter width more accurately reproduces yaw aligned momentum theory at the disk, it also introduces numerical oscillations in the wake flow field which can introduce errors in wake analysis.
However, the rotor-normal induction, when normalized by the yaw aligned induction ($a_n(\gamma)/a_n(\gamma=0)$), shown in Figure~\ref{fig:a_comp_direct}(b) demonstrates that the normalized quantities are less sensitive to the numerical setup.
Therefore, in the results in \S\ref{sec:results}, where analysis of the wake velocity is required (for $\delta u_0$ and $\delta v_0$) and normalized quantities are presented, we use the disk correction with $M$ given by Eq.~\eqref{eq:M_corr} and a larger filter width $\Delta/D=3h/(2D)$.
In \S\ref{sec:results}, where unnormalized quantities are presented and the wake flow is not analyzed, we use a smaller filter width $\Delta/D=0.29h/D=0.032$ which reproduces well-accepted momentum theory for the yaw aligned turbine and does not require the disk velocity correction \cite[]{shapiro2019filtered}.

\section{Glauert induction and power-yaw model}
\label{sec:glauert}

\cite{glauert1926general} developed a model for the relationship between the thrust coefficient $C_T$ and the induction normal to the rotor $a_n^\mathrm{g}$ \cite[see derivation in][]{burton2011wind}
\begin{equation}
C_T = 4 a_n^\mathrm{g} \sqrt{1-a_n^\mathrm{g}(2\cos(\gamma)-a_n^\mathrm{g})}.
\label{eq:ct_glaurt}
\end{equation}
Eq.~\eqref{eq:ct_glaurt} can be solved iteratively for $a_n^\mathrm{g}$ given a known $C_T$ from the initial condition of the yaw aligned induction.
The Glauert model for $C_P$ is
\begin{equation}
C_P^\mathrm{g} = 4 a_n^\mathrm{g} \sqrt{1-a_n^\mathrm{g}(2\cos(\gamma)-a_n^\mathrm{g})}(\cos(\gamma)-a_n^\mathrm{g}),
\label{eq:cp_glauert}
\end{equation}
and the Glauert power ratio model is $P_r^\mathrm{g}(\gamma)=C_P^\mathrm{g}(\gamma)/C_P^\mathrm{g}(\gamma=0)$, where $C_P^\mathrm{g}$ is estimated using Eq.~\eqref{eq:cp_glauert}.

\section{Far-wake model}
\label{sec:far}

The inviscid near-wake model developed in \S\ref{sec:model} can provide the initial conditions for self-similar far-wake models.
The streamwise and spanwise velocity initial conditions are $u_4$ and $v_4$, respectively (see \S\ref{sec:model}).
We use a far-wake model based on the analytical integration of the lifting line model \cite[]{shapiro2018modelling} shown in \citet{howland2019wind}.
The wind turbine wakes are modeled as two-dimensional Gaussian velocity deficits \cite[]{bastankhah2014new, shapiro2018modelling,howland2019wind}.
The model is steady-state and two-dimensional.
We define the upwind turbine with the index $i$ and the downwind turbine with the index $j$.
The velocity deficit associated with an upwind turbine $i$ is
\begin{equation}
du_i(x, y) = \delta u_i(x) \frac{D^2}{8 \sigma_{0,i}^2} \exp{\left(-\frac{(y - y_{c,i}(x))^2}{2 \sigma_{0,i}^2 d_{i}^2(x)}\right)},
\label{eq:du}
\end{equation}
where $D$ is the turbine diameter and the streamwise and spanwise directions are $x$ and $y$, respectively.
The coordinate system is defined with respect to the position of the upwind turbine $i$, such that the centroid of turbine $i$ is at $x=0$ and $y=0$.
The normalized far-wake diameter as a function of the streamwise location $x$ is $d_{i}(x)=1+ k_{w,i} \log \left(1+\exp[2(x/D - 1)]\right)$.
The wake spreading coefficient is $k_w$ and the proportionality constant of the presumed Gaussian wake is $\sigma_0$.
The lateral centroid of the wake of turbine $i$ is $y_{c,i}$.
With freestream wind $u_\infty$ in the $x$-direction and zero freestream wind in the spanwise direction, the streamwise velocity deficit $\delta u_i(x)$ is modeled as \cite[]{shapiro2018modelling}
\begin{equation}
\delta u_i(x) = \frac{u_\infty - u_{4,i}}{d_{i}^2(x)} \frac{1}{2} \left[1+\mathrm{erf}\left(\frac{x}{\sqrt{2}D/2}\right)\right],
\label{eq:delta_uu}
\end{equation}
and the wake centerline lateral velocity as a function of the $x$ position for the upwind turbine is
\begin{equation}
\delta v_i(x) = \frac{-v_{4,i}}{d_{i}^2(x)} \frac{1}{2} \left[1+\mathrm{erf}\left(\frac{x}{\sqrt{2}D/2}\right)\right].
\label{eq:delta_v}
\end{equation}
The lateral centroid of the wake, produced by upwind turbine $i$, is given by
\begin{equation}
y_{c,i}(x) = \int_{x_{0,i}}^{x} \frac{-\delta v_i(x^\prime)}{u_{\infty}} dx^\prime.
\end{equation}
The rotor averaged velocity deficit is \cite[]{howland2019wind}
\begin{equation}
\Delta u_{i,j}(x) = \frac{\sqrt{2 \pi} \delta u_i(x) d_{i}(x) D}{16 \sigma_{0,i}} \left [ \mathrm{erf} \left(\frac{y_T+D/2 - y_{c,i}(x)}{\sqrt{2} \sigma_{0,i} d_{i}(x)}\right) - \mathrm{erf} \left(\frac{y_T-D/2 - y_{c,i}(x)}{\sqrt{2} \sigma_{0,i} d_{i}(x)}\right) \right],
\label{eq:delta_u_ij}
\end{equation}
where the lateral turbine centroid of downwind turbine $j$ is $y_T$.
The rotor averaged velocity at the downwind turbine $j$ is therefore given by
\begin{equation}
u_{e,j} = u_\infty - \Delta u_{i,j},
\end{equation}
and the power production of turbine $j$, following Eq.~\eqref{eq:power_ADM}, is
\begin{equation}
P_j = \frac{1}{2} \rho C_{T,j}^\prime A_d \left[(1-a_{n,j}) \cos(\gamma_j) u_{e,j} \right]^3.
\end{equation}
The power of turbine $i$ is modeled similarly.

\FloatBarrier

\bibliographystyle{jfm}

\end{document}